\documentclass[MSNbibl,nameyear,seceqn,dvips]{arxstspdf}
\usepackage{flushend}
\usepackage{stfloats}
\volume{28}
\issue{1}
\pubyear{2013}
\firstpage{40}
\lastpage{68}
\doi{10.1214/12-STS395} 

\makeatletter
\newproclaim{Remark}{Remark}
\makeatother

\begin{document}
\begin{frontmatter}

\title{New Important Developments in Small Area Estimation}
\runtitle{New Important Developments in Small Area Estimation}

\begin{aug}
\author[a]{\fnms{Danny} \snm{Pfeffermann}\corref{}\ead[label=e1]{msdanny@soton.ac.uk}}
\runauthor{D. Pfeffermann}

\affiliation{University of Southampton and Hebrew University of Jerusalem}

\address[a]{Danny Pfeffermann is Professor of Statistics, Southampton Statistical Sciences Research Institute, University of
Southampton, Southampton, SO17 1BJ, United Kingdom, and Department of Statistics, Hebrew University of Jerusalem, Jerusalem,
91905, Israel \printead{e1}.}

\end{aug}

\begin{abstract}
The problem of small area estimation (SAE) is how to produce reliable
estimates of characteristics of interest such as means, counts,
quantiles, etc., for areas or domains for which only small samples or no
samples are available, and how to assess their precision. The purpose of
this paper is to review and discuss some of the new important
developments in small area estimation methods. Rao [\textit{Small Area Estimation} (2003)] wrote a very
comprehensive book, which covers all the main developments in this topic
until that time. A few review papers have been written after 2003, but
they are limited in scope. Hence, the focus of this review is on new
developments in the last 7--8 years, but to make the review more
self-contained, I also mention shortly some of the older developments.
The review covers both design-based and model-dependent methods, with
the latter methods further classified into frequentist and Bayesian
methods. The style of the paper is similar to the style of my previous
review on SAE published in 2002, explaining the new problems
investigated and describing the proposed solutions, but without dwelling
on theoretical details, which can be found in the original articles. I
hope that this paper will be useful both to researchers who like to
learn more on the research carried out in SAE and to practitioners who
might be interested in the application of the new methods.
\end{abstract}

\begin{keyword}
\kwd{Benchmarking}
\kwd{calibration}
\kwd{design-based methods}
\kwd{empirical likelihood}
\kwd{informative sampling}
\kwd{matching priors}
\kwd{measurement errors}
\kwd{model checking}
\kwd{M-quantile}
\kwd{ordered means}
\kwd{outliers}
\kwd{poverty mapping}
\kwd{prediction intervals}
\kwd{prediction MSE}
\kwd{spline regression}
\kwd{two-part model}.
\end{keyword}

\end{frontmatter}

\section{Preface}

The problem of small area estimation (SAE) is how to produce reliable
estimates of characteristics of interest such as means, counts,
quantiles, et cetera, for areas or domains for which only small samples
or no samples are available. Although the point estimators are usually
of first priority, a related problem is how to assess the estimation
(prediction) error.

The great importance of SAE stems from the fact that many new programs,
such as fund allocation for needed areas, new educational or health
programs and environmental planning rely heavily on these estimates. SAE
techniques are also used in many countries to test and adjust the counts
obtained from censuses that use administrative records.

In 2002 I published a review paper with a similar title (\cite{Pf02}). In that year small area estimation (SAE) was flourishing both in
research and applications, but my own feeling then was that the topic
has been more or less exhausted in terms of research and that it will
just turn into a routine application in sample survey practice. As the
past 9 years show, I was completely wrong; not only is the research in
this area accelerating, but it now involves some of the best known
statisticians, who otherwise are not involved in survey sampling theory
or applications. The diversity of new problems investigated is
overwhelming, and the solutions proposed are not only elegant and
innovative, but also very practical.

\citet{Ra03} published a comprehensive book on SAE that covers all the
main developments in this topic up to that time. The book was written
about ten years after the review paper of Ghosh and Rao (\citeyear{GhRa94}),
published in \textit{Statistical Science}, which stimulated much of the
early research in SAE. Since 2003, a few other review papers have been
published; see, for example, Rao (\citeyear{Ra05}, \citeyear{Ra08}), Jiang and Lahiri (\citeyear{JiLa06a}, \citeyear{JiLa06b}),
Datta (\citeyear{Da09}) and Lehtonen and Veiganen (\citeyear{LeVe09}). Notwithstanding, SAE is
researched and applied so broadly that I decided that the time is ripe for a new
comprehensive review that focuses on the main developments in the last
7--8 years that I am aware of, and which are hardly covered in the review
papers mentioned above. The style of the paper is similar to the style
of my previous review, explaining the problems investigated and
describing the proposed solutions, but without dwelling on theoretical
details, which can be found in the original articles. For further
clarity and to make the paper more self-contained, I~start with a short
background and overview some of the ``older'' developments. I hope
that this paper will be useful to researchers who wish to learn about
the research carried out in SAE and to practitioners who might be
interested in applying the new methods.

\section{Some Background}

The term ``SAE'' is somewhat confusing, since it is the size of
the sample in the area that causes estimation problems, and not the size
of the area. Also, the ``areas'' are not necessarily geographical
districts and may define another grouping, such as socio-demographic
groups or types of industry, in which case they are often referred to as
domains. Closely related concepts in common use are ``poverty
mapping'' or ``disease mapping,'' which amount to SAE of poverty
measures or disease incidence and then presenting the results on a map,
with different colors defining different levels (categories) of the
estimators. What is common to most small area estimation problems is
that point estimators and error measures are required for every area
separately, and not just as an average over all the areas under
consideration.

SAE methods can be divided broadly into ``design-based'' and
``model-based'' methods. The latter methods use either the
frequentist approach or the full Bayesian methodology, and in some cases
combine the two, known in the SAE literature as ``empirical
Bayes.'' Design-based methods often use a model for the construction of
the estimators (known as ``model assisted''), but the bias,
variance and other properties of the estimators are evaluated under the
randomization (design-based) distribution. The randomization
distribution of an estimator is the distribution over all possible samples that could be
selected from the target population of interest under the sampling
design used to select the sample, with the population measurements
considered as fixed values (parameters). Model-based methods on the
other hand usually condition on the selected sample, and the inference
is with respect to the underlying model.

A common feature to design- and model-based SAE is the use of auxiliary
covariate information, as obtained from large surveys and/or
administrative records such as censuses and registers. Some estimators
only require knowledge of the covariates for the sampled units and the
true area means of these covariates. Other estimators require knowledge
of the covariates for every unit in the population. The use of auxiliary
information for SAE is vital because with the small sample sizes often
encountered in practice, even the most elaborated model can be of little
help if it does not involve a set of covariates with good predictive power for
the small area quantities of interest.\vspace*{-1pt}

\section{Notation}

Consider a population $U$ of size $N$, divided into $M$ exclusive and
exhaustive areas $U_{1} \cup\cdots \cup U_{M}$ with $N_{i}$ units in area
$i$, $\sum_{i = 1}^{M} N_{i} = N$. Suppose that samples are available for
$m \le M$ of the areas, and let $s = s_{1} \cup\cdots \cup s_{m}$ define
the overall sample, where $s_{i}$ of size $n_{i}$ is the sample observed
for sampled area $i$, $\sum_{i = 1}^{m} n_{i} = n$. Note that $n_{i}$ is random
unless a planned sample of fixed size is taken in that area. Let $y$
define the characteristic of interest, and denote by $y_{ij}$ the
response value for unit $j$ belonging to area $i$, $i = 1,\ldots,M $, $j =
1,\ldots,N_{i}$ with sample means $\bar{y}_{i} = \sum_{j = 1}^{n_{i}}
y_{ij}/n_{i}$, where we assume\vspace*{2pt} without loss of generality that the
sample consists of the first $n_{i}$ units. We denote by
$\mathrm{x}_{ij} = (x_{1ij},\ldots,x_{pij})'$ the covariate values
associated with unit $(i,j)$ and by $\mathrm{\bar{x}}_{i} = \sum_{j =
1}^{n_{i}} \mathrm{x}_{ij} /n_{i}$ the column vector of sample means.
The corresponding vector of true area means is $\bar{X}_{i} = \sum_{j =
1}^{N_{i}} \mathrm{x}_{ij} /N_{i}$. The area target quantity is denoted
by $\theta_{i}$; for example, $\theta_{i} = \bar{Y}_{i} = \sum_{j =
1}^{N_{i}} y_{ij}/N_{i}$, the response area mean. Estimating a
proportion is a special case where $y_{ij}$ is binary. In other
applications $\theta_{i}$ may represent a count or a quantile.

\section{Design-Based Methods}\label{sec4}

\subsection{Design-Based Estimators in Common Use}\label{sec4.1}

A recent comprehensive review of design-based methods in SAE is provided
by Lehtonen and Veijanen (\citeyear{LeVe09}). Here I only overview some of the basic
ideas. Suppose that the sample is selected by simple random sampling
without replacement (SRSWOR) and that the target quantities of interest
are the means~$\bar{Y}_{i}$. Estimation of a mean contains as special
cases the estimation of a proportion and the estimation of the area
distribution $F_{i}(t) = \sum_{j \in U_{i}} v_{ij}/N_{i}$, in which case
$v_{ij} = I(y_{ij} \le t)$, where $I(A)$ is the indicator function.
Estimators of the percentiles of the area distribution are commonly
obtained from the estimated distribution.

If no covariates are available the \textit{direct} design-unbiased
estimator of the area mean and its conditional design variance over the
\textit{randomization} distribution for given $n_{i}$ are given by
\begin{eqnarray}\label{4.1}
\bar{y}_{i} &=& \sum_{j = 1}^{n_{i}} y_{ij} /n_{i} ;\nonumber\\[-8pt]\\[-8pt]
V_{D}[\bar{y}_{i}|n_{i}] &=& (S_{i}^{2}/n_{i})[1 - (n_{i}/N_{i})],\nonumber
\end{eqnarray}
where $S_{i}^{2} = \sum_{j = 1}^{N_{i}} (y_{ij} -
\bar{Y}_{i})^{2}/(N_{i} - 1)$. The term ``direct'' is used to
signify an estimator that only uses the data available for the target
area at the specific time of interest. The variance
$V_{D}[\bar{y}_{i}|n_{i}]$ is $O(1/n_{i})$, and for small $n_{i}$ it is
usually large, unless $S_{i}^{2}$ is sufficiently small.

Next suppose that covariates $\mathrm{x}_{ij}$ are also observed with
$\mathrm{x}_{1ij} \equiv 1$. An estimator in common\vadjust{\goodbreak} use that utilizes the
covariate information is the \textit{synthetic} estimator,
\begin{eqnarray}\label{4.2}
\hat{\bar{Y}}_{\mathrm{reg},i}^{\mathrm{syn}} = \bar{X}'_{i}\hat{B} =
\frac{1}{N_{i}}\sum_{j = 1}^{N_{i}} (\mathrm{x}'_{ij} \hat{B}),
\end{eqnarray}
where $\hat{B} = [\sum_{i = 1}^{m} \sum_{j = 1}^{n_{i}} \mathrm{x}_{ij}\mathrm{x}'_{ij}]
^{ - 1}\sum_{i = 1}^{m} \sum_{j = 1}^{n_{i}} \mathrm{x}_{ij}y_{ij}$ is the
ordinary least square (OLS) estimator. Notice that under SRSWOR,
$\hat{B}$ is approximately design-unbiased and consistent for the vector
$B$ of regression coefficients computed from all the population values,
irrespective of whether a linear relationship \mbox{between} $y$ and~$\mathrm{x}$ exists in the population. The design-unbiasedness and
consistency are with respect to the randomization distribution, letting
$N$ and $n$ increase to infinity in a proper way. An estimator is
approximately design-unbiased if the randomization bias tends to zero as
the sample size increases. The term ``synthetic'' refers to the
fact that an (approximately) design-unbiased estimator computed from all
the areas ($\hat{B}$ in the present case) is used for every area
separately, assuming that the areas are ``homogeneous'' with
respect to the quantity being estimated. Thus, synthetic estimators
borrow information from other ``similar areas'' and they are
therefore \textit{indirect} estimators.

The obvious advantage of the synthetic estimator over the simple sample
mean or other direct estimators such as the regression estimator
$\hat{\bar{Y}}^{\mathrm{dir}}_{\mathrm{reg},i} = \bar{y}_{i} + (\bar{X}_{i} -
\bar{x}_{i})'\hat{B}_{i}$, where $\hat{B}_{i}$ is computed only from the
data observed for area $i$, is that
$\operatorname{Var}_{D}(\hat{\bar{Y}}_{\mathrm{reg},i}^{\mathrm{syn}}) = O(1/n)$, and $n = \sum_{i =
1}^{m} n_{i}$ is usually large. The use of the synthetic estimator is
motivated (``assisted'') by a linear regression model of $y$ on
$\mathrm{x}$ in the population with a common vector of coefficients.
However, for $\mathrm{x}_{1ij} \equiv 1$,
$E_{D}(\hat{\bar{Y}}_{\mathrm{reg},i}^{\mathrm{syn}} - \bar{Y}_{i}) \cong -
\bar{X}'_{i}(B_{i} - B)$, where $B_{i}$ is the OLS computed from all the
population values in area $i$. Thus, if in fact different regression
coefficients $B_{i}$ operate in different areas, the synthetic estimator
may have a large bias. When the sample is selected with unequal
probabilities, the OLS estimator $\hat{B}$ in (\ref{4.2}) is commonly replaced
by the probability weighted (PW) estimator
\[
\hat{B}_{\mathrm{pw}} = \Biggl[\sum_{i =
1}^{m} \sum_{j = 1}^{n_{i}} w_{ij}\mathrm{x}_{ij}\mathrm{x}'_{ij}\Biggr] ^{ - 1}\sum_{i = 1}^{m}
\sum_{j = 1}^{n_{i}} w_{ij}\mathrm{x}_{ij}y_{ij},
\]
 where $\{ w_{ij} = 1/\Pr
[(i,j) \in s]\}$ are the base sampling weights.

In order to deal with the possible large bias of the synthetic
estimator, it is common\vadjust{\goodbreak} to estimate the bias and then subtract it from
the synthetic estimator. The resulting \textit{survey regression}
estimator takes the form
\begin{eqnarray}\label{4.3}
\hat{\bar{Y}}_{i}^{\mathrm{S\mbox{--}R}} &=& \bar{X}'_{i}\hat{B}_{\mathrm{pw}} +
\frac{1}{N_{i}}\sum_{j = 1}^{n_{i}} w_{ij}(y_{ij} - \mathrm{x}'_{ij}\hat{B}_{\mathrm{pw}})\nonumber\\[-8pt]\\[-8pt]
&=& \hat{\bar{Y}}_{i,\mathrm{H\mbox{--}T}} + (\bar{X}_{i} - \hat{\bar{X}}_{i,\mathrm{H\mbox{--}T}})'\hat{B}_{\mathrm{pw}},\nonumber
\end{eqnarray}
where $(\hat{\bar{Y}}_{i,\mathrm{H\mbox{--}T}},\hat{\bar{X}}_{i,\mathrm{H\mbox{--}T}})$ are the
Horvitz--Thompson (H--T) estimators of $(\bar{Y}_{i},\bar{X}_{i})$. The
estimator (\ref{4.3}) is approximately design-unbiased and performs well when
the covariates have good predictive power, but the variance is back to
order $O(1/n_{i})$. The variance is often reduced by multiplying the
bias correction $\sum_{j = 1}^{n_{i}} w_{ij}(y_{ij} -
\mathrm{x}'_{ij}\hat{B}_{\mathrm{pw}})/N_{i}$ by $N_{i}/\sum_{j = 1}^{n_{i}} w_{ij} =
N_{i}/\hat{N}_{i}$.

A compromise between the possibly large bias of the synthetic estimator
and the possibly large variance of the survey regression estimator is
achieved by taking a linear combination of the two. The resulting
\textit{combined} (\textit{composite}) estimator is defined as
\begin{equation}\label{4.4}
\hspace*{15pt}\hat{\bar{Y}}_{i}^{\mathrm{COM}} = \delta_{i}\hat{\bar{Y}}_{i}^{\mathrm{S\mbox{--}R}} +
(1 - \delta_{i})\hat{\bar{Y}}_{\mathrm{reg},i}^{\mathrm{syn}};\quad 0 \le\delta_{i} \le 1.\hspace*{-15pt}
\end{equation}
Ideally, the coefficient $\delta_{i}$ should be chosen to minimize the
mean square error (MSE) of $\hat{\bar{Y}}_{i}^{\mathrm{COM}}$, but assessing
sufficiently accurately the bias of the synthetic estimator for a given area is
usually impossible. Hence, it is common to let $\delta_{i}$ depend on
the sample size $n_{i}$ in the area, such that the larger $n_{i}$, the
larger is $\delta_{i}$. See \citet{Ra03} for review of other combined estimators, and methods of
specifying $\delta_{i}$.

\subsection{Some New Developments in Design-Based Small Area Estimation}

A general class of estimators is obtained by calibrating the base
sampling weights $w_{ij}$. Suppose that the population can be
partitioned into $C$ calibration groups $U = U_{(1)} \cup\cdots \cup
U_{(C)}$ with known totals $t_{\mathrm{x}(c)}$ of the auxiliary
variables in the groups, such that each area $U_{i}$ belongs to one of
the groups. Let $s = s_{(1)} \cup\cdots \cup s_{(C)}$ define the respective
partitioning of the sample. In a special case $C = 1$ and $U_{(1)} = U$.
The \textit{calibrated} estimator of the mean $\bar{Y}_{i}$ is computed
as
\begin{equation}\label{4.5}
\hspace*{12pt}\hat{\bar{Y}}_{i}^{\mathrm{cal}} = \sum_{j = 1}^{n_{i}} w_{ij}^{c}
y_{ij}/N_{i} ;\quad \sum_{i,j \in s_{(c)}} w_{ij}^{c} \mathrm{x}_{ij} =
t_{\mathrm{x}(c)}.\hspace*{-12pt}
\end{equation}
The calibration weights $\{ w_{ij}^{c}\}$ are chosen so that they
minimize an appropriate distance from the base weights $\{ w_{ij}\}$,
subject\vadjust{\goodbreak} to satisfying the constraints $\sum_{i,j \in s_{(c)}} w_{ij}^{c}
\mathrm{x}_{ij} = t_{\mathrm{x}(c)}$. For example, when using the
distance $\chi^{2} = \sum_{i,j \in s_{(c)}} (w_{ij}^{c} -
w_{ij})^{2}/w_{ij}$ and $x_{1ij} \equiv 1$, the calibrated weights are
\begin{eqnarray}\label{4.6}
w_{ij}^{c} &=& w_{ij}g_{ij};\nonumber\\
 g_{ij} &=& \biggl\{ 1 + \bigl(t_{\mathrm{x}(c)} -
\hat{t}_{\mathrm{x}(c),\mathrm{H\mbox{--}T}}\bigr)'\\
&&\hspace*{24pt}{}\cdot\biggl[\sum_{i,j \in s_{(c)}}
w_{ij}\mathrm{x}_{ij}\mathrm{x}'_{ij}\biggr]^{ - 1} \mathrm{x}_{ij}\biggr\},\nonumber
\end{eqnarray}
where $\hat{t}_{\mathrm{x}(c),\mathrm{H\mbox{--}T}}$ is the H--T estimator of the total
$t_{\mathrm{x}(c)}$. When $U_{c} = U_{i}$ (the calibration group is the
domain), $\hat{\bar{Y}}_{i}^{\mathrm{cal}}$ is the familiar \textit{generalized
regression} (GREG) estimator in the domain.

Calibration of the sampling weights is in broad use in sample survey
practice, not only for SAE. See Kott (\citeyear{Ko09}) for a recent comprehensive
review and discussion. The rationale of the use of calibrated estimators
in SAE is that if $y$ is approximately a linear combination of
$\mathrm{x}$ in $U_{(c)}$, then $\bar{Y}_{i}
\cong\bar{X}'_{i}B_{(c)}$ for domains $i \in U_{c}$, and since\vspace*{-2pt}
$\sum_{i,j \in s_{(c)}} w_{ij}^{c} \mathrm{x}_{ij} = t_{\mathrm{x}(c)}$,
$\hat{\bar{Y}}_{i}^{\mathrm{cal}} = \sum_{j = 1}^{n_{i}} w_{ij}^{c} y_{ij}/N_{i}$
is expected to be a good estimator of $\bar{Y}_{i}$. Indeed, the
advantage of estimator (\ref{4.5}) over (\ref{4.2}) is that it is assisted by a
model that only assumes common regression coefficients within the groups
$U_{(c)}$, and not for all the domains, as implicitly assumed by
estimator (\ref{4.2}). Estimator (\ref{4.5}) is approximately design-unbiased
irrespective of any model, but $\operatorname{Var}_{D}(\hat{\bar{Y}}_{i}^{\mathrm{cal}}|n_{i}) =
O(1/n_{i})$, which may still be large.

Another way of calibrating the weights is by use of \textit{instrumental
variables} (Estevao and S\"{a}rndal, \citeyear{EsSa04a}, \citeyear{EsSa04b}). Denote the vector of
instrument values for unit $(i,j)$ by $h_{ij}$. The calibrated weights
are defined as
\begin{eqnarray}\label{4.7}
\hspace*{30pt}w_{ij}^{\mathrm{ins}} &=& w_{ij}(1 + g'_{c}h_{ij}) ;\hspace*{-30pt}\nonumber\\[-8pt]\\[-8pt]
 g'_{c} &=&
\bigl(t_{\mathrm{x}(c)} - \hat{t}_{\mathrm{x(c),\mathrm{H\mbox{--}T}}}\bigr)'\biggl[\sum_{i,j \in
s_{(c)}} w_{ij}h_{ij}\mathrm{x}'_{ij}\biggr]^{ - 1}.\nonumber
\end{eqnarray}
Note that the instrument values need only be\break known for the sampled units
in $s_{(c)}$ and that\break $\sum_{i,j \in s_{(c)}} w_{ij}^{\mathrm{ins}}
\mathrm{x}_{ij} = t_{c\mathrm{x}}$, thus satisfying the same constraints
as before. The calibrated estimator of $\bar{Y}_{i}$ is now
$\hat{\bar{Y}}_{i,\mathrm{ins}}^{\mathrm{cal}} = \sum_{j = 1}^{n_{i}}
w_{ij}^{\mathrm{ins}}\cdot\allowbreak
y_{ij}/N_{i}$. When $h = \mathrm{x}, w_{ij}^{\mathrm{ins}} = w_{ij}^{c}$. The use
of\vspace*{1pt} instruments replaces the search for an appropriate distance function
by imposing a structure on the calibration weights, and it allows one,
in principle, to find the best instruments in terms of minimizing an
approximation to the variance of the calibrated estimator. However, as
noted by Estevao and S\"{a}rndal (\citeyear{EsSa04b}), the resulting optimal weights depend on
unknown population quantities which, when estimated from the sample, may
yield unstable estimators. See Kott (\citeyear{Ko09}) for further discussion.

The synthetic estimator (\ref{4.2}), the survey regression estimator (\ref{4.3}) and
the various calibrated estimators considered above are all assisted by
models that assume a linear relationship between $y$ and
$\mathrm{x}$. These estimators only require knowledge of the covariates
for the sampled units, and the area (or group) totals of these
covariates. Lehtonen, S{\"a}rndal and Veijanen (\citeyear{LeSaVe03}, \citeyear{LeSaVe05}) consider the use of generalized
linear models (GLM), or even generalized linear mixed models (GLMM) as
the assisting models, which require knowledge of the covariates for
every element in the population. Suppose that $E_{M}(y_{ij}) =
f(\mathrm{x}_{ij};\psi)$ for some nonlinear function $f( \cdot)$ with an
unknown vector parameter $\psi$, where $E_{M}( \cdot)$ defines the
expectation under the model. A simple important example is where
$f(\mathrm{x}_{ij};\psi)$ is the logistic function. Estimating $\psi$ by
the \textit{pseudo-likelihood} (PL) approach yields the estimator
$\hat{\psi} _{\mathrm{pl}}$ and predicted values $\{ \hat{y}_{ij} =
f(\mathrm{x}_{ij};\hat{\psi} _{\mathrm{pl}})\}$. The PL approach consists of
estimating the likelihood equations that would be obtained in case of a
census by the corresponding H--T estimators (or weighting each score
function by its sampling weight), and then solving the resulting
estimated equations. The synthetic and ``generalized GREG'' estimators
are computed as
\begin{eqnarray}\label{4.8}
\qquad\hat{\bar{Y}}_{\mathrm{GLM},i}^{\mathrm{syn}} &=& \frac{1}{N_{i}}\sum_{j = 1}^{N_{i}}
f(\mathrm{x}_{ij};\hat{\psi} _{\mathrm{pl}});\nonumber\\
 \hat{\bar{Y}}_{\mathrm{GLM},i}^{\mathrm{GREG}} &=&
\hat{\bar{Y}}_{\mathrm{GLM},i}^{\mathrm{syn}}\\
&&{} + \frac{1}{N_{i}}\sum_{j = 1}^{n_{i}}
w_{ij}[y_{ij} - f(\mathrm{x}_{ij};\hat{\psi} _{\mathrm{pl}})].\nonumber
\end{eqnarray}

A further extension is to include random area effects in the assisting
model, assuming $E_{M}(y_{ij}|\mathrm{x}_{ij},\break u_{i}) =
f(\mathrm{x}_{ij},u_{i};\psi^*)$, $E_{M}(u_{i}) = 0$, $\operatorname{Var}_{M}(u_{i}) =
\sigma_{u}^{2}$. Estimation of the fixed parameters $\psi
^*$, $\sigma_{u}^{2}$ and the random effects $u_{i}$ is now under the model,
ignoring the sampling weights. The extended synthetic and generalized
GREG estimators are defined similarly to (\ref{4.8}), but with
$f(\mathrm{x}_{ij};\hat{\psi} _{\mathrm{pl}})$ replaced by
$f(\mathrm{x}_{ij},\hat{u}_{i};\hat{\psi}^*)$. For sufficiently large
sample size $n_{i}$, the extended generalized GREG is approximately
design-unbiased for the true area mean,\vadjust{\goodbreak} but it is not clear how to
estimate the design (randomization) variance in this case in a way that
accounts for the prediction of the random effects. \citet{ToRa08}
compare the MSE of model-based predictors and a GREG assisted by a
linear mixed model (LMM).

\citet{JiLa06a} propose the use of model-dependent estimators
that are design-consistent under the randomization distribution as the
area sample sizes increase. The basic idea is to model the direct\vspace*{1pt}
estimators $\hat{\bar{Y}}_{iw} = \sum_{j = 1}^{n_{i}} w_{ij}y_{ij}/
\sum_{j = 1}^{n_{i}} w_{ij}$ instead of the individual observations
$y_{ij}$, and then employ the empirical best predictor of the area mean
under the model. The authors consider the general two-level model
$E_{M}[\hat{\bar{Y}}_{iw}|u_{i}] = \xi_{i} =
\xi(u_{i},\hat{\bar{X}}_{iw};\psi)$,\break where the $u_{i}$s are independent
random area effects with zero mean and variance $\sigma_{u}^{2}$,
$\hat{\bar{X}}_{iw} = \sum_{j = 1}^{n_{i}} w_{ij}\mathrm{x}_{ij}/ \break\sum_{j
= 1}^{n_{i}} w_{ij}$, and $\xi( \cdot)$ is some known function with
unknown parameters $\psi$. The empirical best predictor is the best
predictor under the model (minimum expected quadratic loss), but with
the parameters $\psi$ replaced by model consistent estimators;
$\hat{\bar{Y}}_{i}^{\mathrm{EBP}} = E_{M}(\xi_{i}|\hat{\bar{Y}}_{iw},\hat{\bar{X}}_{iw};\hat{\psi}
)$. The estimator is shown to be model-consistent under correct model
specification and design-consistent for large $n_{i}$, even if the model
is misspecified, thus robustifying the estimation. The authors develop
estimators of the prediction mean squared error (PMSE) for bounded
sample sizes~$n_i$, with bias of desired order $o(1/m)$, where $m$ is the
number of sampled areas. The PMSE is computed with respect to the model
holding for the individual observations and over the randomization
distribution. The use of design consistent estimators in SAE is somewhat
questionable because of the small sample sizes in some or all of the
areas, but it is nonetheless a desirable property. This is so because it
is often the case that in some of the areas the samples are large, and
it is essential that an estimator should work well at least in these
areas, even if the model fails. Estimators with large randomization bias
even for large samples do not appeal to practitioners.

\citet{ChCh09} propose the use of model-based direct
estimators (MBDE). The idea is to fit a model for the population values,
compute the weights defining the Empirical Best Linear Unbiased
Predictor (EBLUP) of the population total under the model and then use
the weights associated with a given area to compute an almost direct
estimator. The model fitted for the population values $Y_{U}$ is the
general linear model,
\begin{eqnarray}\label{4.9}
Y_{U} &=& X_{U}\beta + \varepsilon_{U};\quad E(\varepsilon_{U}) = 0,\nonumber\\[-8pt]\\[-8pt]
E(\varepsilon_{U}\varepsilon '_{U}) &=& \Sigma = \left[ \matrix{
\Sigma_{ss} & \Sigma_{sr} \cr \Sigma_{rs} & \Sigma_{rr}} \right],\nonumber
\end{eqnarray}
where $s$ signifies the sample of size $n$, and $r$ signifies the
sample-complement of size $(N - n)$. As seen later, the models in common
use for SAE defined by (\ref{5.1}) and (\ref{5.3}) below are special cases of (\ref{4.9}).
Let $y_{s}$ denote the column vector of sample outcomes. For known
$\Sigma$, the BLUP of the population total $t_{y} = \sum_{k = 1}^{N}
y_{k}$ under the model is
\begin{eqnarray}\label{4.10}
\hat{t}_{y}^{\mathrm{BLUP}} &=& 1'_{n}y_{s} + 1'_{N - n}[X_{r}\hat{\beta}
_{\mathrm{GLS}}\nonumber\hspace*{-20pt}\\
&&\hspace*{60pt}{} + \Sigma_{rs}\Sigma_{ss}^{ - 1}(y_{s} - X_{s}\hat{\beta} _{\mathrm{GLS}})]\hspace*{-20pt}\\
&=& \sum_{k \in s} w_{k}^{\mathrm{BLUP}}y_{k},\nonumber\hspace*{-20pt}
\end{eqnarray}
where $1'_k$ is a row vector of ones of length $k$, $X_s(X_r)$ is the design matrix corresponding to the sampled (nonsampled) units and
$\hat{\beta} _{\mathrm{GLS}}$ is the generalized least square estimator.
The EBLUP is $\hat{t}_{y}^{\mathrm{EBLUP}} = \break\sum_{k \in s} w_{k}^{\mathrm{EBLUP}}y_{k}$,
where the EBLUP weights are the same as in (\ref{4.10}), but with estimated
parameters. The MBDE of the true mean in area $i$ is
\begin{equation}\label{4.11}\qquad
\hat{\bar{Y}}_{i}^{\mathrm{MBD}} = \sum_{j \in s_{i}} w_{j}^{\mathrm{EBLUP}}y_{j}\Big/
\sum_{j \in s_{i}} w_{j}^{\mathrm{EBLUP}}.
\end{equation}
The authors derive estimators for the bias and variance of the MBDE and
illustrate its robustness to certain model misspecifications. Note,
however, that $\hat{\bar{Y}}_{i}^{\mathrm{MBD}}$ is a ratio estimator and
therefore may have a nonnegligible bias in areas $i$ with small sample
size.

All the estimators considered so far assume a given sampling design with
random area sample sizes.\break When the target areas are known in advance,
considerable gains in efficiency can be achieved by modifying the
sampling design and in particular, by controlling the sample sizes
within these areas. In a recent article, Falrosi and Righi (\citeyear{FaRi08})
propose a general strategy for multivariate multi-domain estimation that
guarantees that the sampling errors of the domain estimators are lower
than pre-specified thresholds. The strategy combines the use of a
balanced sampling technique and GREG estimation, but extensions to the
use of synthetic estimators and model-based estimation are also
considered. A~successful application of this strategy requires good
predictions of weighted sums of residuals featuring in the
variance
expressions, and it may happen that the resulting overall\vadjust{\goodbreak} sample size is
far too large, but this is a promising approach that should be studied
further.

\subsection{Pros and Cons of Design-Based Small Area Estimation}\label{sec4.3}

The apparent advantage of design-based methods is that the estimation is
less dependent on an assumed model, although models are used (assisted)
for the construction of the estimators. The estimators are approximately
unbiased and consistent under the randomization distribution for large
sample sizes within the areas, which as discussed before is a desirable
property that protects against possible model misspecification at least
in large areas.

Against this advantage stand many disadvantages. Direct estimators
generally have large variance due to small sample sizes. The survey
regression estimator is approximately unbiased but may likewise be too
variable. Synthetic estimators have small variance but are generally
biased. Composite estimators have smaller bias than synthetic estimators
but larger variance, and it is not obvious how to best choose the
weights attached to the synthetic estimator and the unbiased estimator. Computation of randomization-based confidence
intervals generally requires large sample normality assumptions, but the
sample sizes in at least some of the areas may be too small to justify
asymptotic normality.

Another limitation of design-based inference (not restricted to SAE) is
that it does not lend itself to conditional inference, for example,
conditioning on the sampled values of the covariates or the sampled
clusters in a two-stage sampling design. This again inflates the
variance of the estimators. Conditional inference is in the heart of
classical statistical inference under both the frequentist and the
Bayesian approaches. Last, but not least, an important limitation of
design-based SAE is that there is no founded theory for estimation in
areas with no samples. The use of the randomization distribution does
not extend to prediction problems, such as the prediction of small area
means for areas with no samples. It is often the case that samples are
available for only a minority of the areas, but estimators and MSE
estimators are required for each of the areas, whether sampled or not.

\section{Model-Based Methods}\label{sec5}

\subsection{General Formulation}\label{sec5.1}

Model-based methods assume a model for the sample data and use the
optimal or approximately optimal predictor\vadjust{\goodbreak} of the area characteristic of
interest under the model. The MSE of the prediction error is likewise
defined and estimated with respect to the model. Note that I now use the
term ``prediction'' rather than estimation because the target
characteristics are generally random under the model. The use of models
overcomes the problems underlying the use of design-based methods, but
it is important to emphasize again that even the most elaborated model
cannot produce sufficiently accurate predictors when the area sample
size is too small, and no covariates with good predictive power are
available. The use of models raises the question of the robustness of
the inference to possible model misspecification, and Sections~\ref{sec6.3}--\ref{sec6.6}
review studies that deal with this problem from different perspectives.
Section~\ref{sec8} considers model selection and diagnostic checking.

Denote by $\theta_{i}$ the target quantity in area $i$ (mean,
proportion, $\dots$). Let $y_{i}$ define the observed responses for area
$i$ and $\mathrm{x}_{i}$ define the corresponding values of the
covariates (when available). As becomes evident below, $y_{i}$ is either
a scalar, in which case $\mathrm{x}_{i}$ is a vector, or $y_{i}$ is a
vector, in which case $\mathrm{x}_{i}$ is usually a matrix. A typical
small area model consists of two parts: The first part models the
distribution (or just the moments) of $y_{i}|\theta_{i};\psi_{(1)}$. The
second part models the distribution (moments) of
$\theta_{i}|\mathrm{x}_{i};\psi_{(2)}$, linking the $\theta_{i}$s to
known covariates and to each other. This is achieved by including in the
model random effects that account for the variability of the
$\theta_{i}$s not explained by the covariates. The hyper-parameters
$\psi = (\psi_{(1)},\psi_{(2)})$ are typically unknown and are estimated
either under the frequentist approach, or under the Bayesian approach
by setting appropriate prior distributions. In some applications the index $i$
may define time, in which case the model for
$\theta_{i}|\mathrm{x}_{i};\psi_{2}$ is a time series model.

\subsection{Models in Common Use}\label{sec5.2}

In this section, I review briefly three models in common use, as most of
the recent developments in SAE relate to these models or extensions of
them. For more details see \citet{Ra03}, Jiang and Lahiri (\citeyear{JiLa06a}, \citeyear{JiLa06b}), \citet{Da09}
and the references\break therein. I assume that the model holding for
the sample data is the same as the model holding in the population, so
that there is no sample selection bias. The case of informative selection of the areas to be sampled or
informative sampling within the selected areas, whereby the sample selection or
response probabilities are related to the response variable even after
conditioning on the model covariates is considered in Section~\ref{sec7}. Notice
that in this case the sample model differs from the population model.

\subsubsection{Area level model}

This model is in broad use when the covariate information is only at the
area level, so that $\mathrm{x}_{i}$ is a vector of known area
characteristics. The model, studied originally for SAE by \citet{FaHe79} is defined as
\begin{equation}\label{5.1}
\tilde{y}_{i} = \theta_{i} + e_{i} ;\quad \theta_{i} =
\mathrm{x}'_{i}\beta + u_{i},
\end{equation}
where $\tilde{y}_{i}$ denotes the direct sample estimator of $\theta_{i}$
(e.g., the sample mean $\bar{y}_{i}$ when the sample is selected by
SRS), and $e_{i}$ represents the sampling error, assumed to have zero
mean and known design (randomization) variance, $\operatorname{Var}_{D}(e_{i}) =
\sigma_{Di}^{2}$. The random effects $u_{i}$ are assumed to be
independent with zero mean and variance $\sigma_{u}^{2}$. For known
$\sigma_{u}^{2}$, the best linear unbiased predictor (BLUP) of
$\theta_{i}$ under this model is
\begin{eqnarray}\label{5.2}
\hat{\theta} _{i} &=& \gamma_{i}\tilde{y}_{i} + (1 -
\gamma_{i})\mathrm{x}'_{i}\hat{\beta} _{\mathrm{GLS}}\nonumber\\
& =&
\mathrm{x}'_{i}\hat{\beta} _{\mathrm{GLS}} + \gamma_{i}(\tilde{y}_{i} -
\mathrm{x}'_{i}\hat{\beta} _{\mathrm{GLS}})\\
& =& \mathrm{x}'_{i}\hat{\beta} _{\mathrm{GLS}} +
\hat{u}_{i}.\nonumber
\end{eqnarray}
The BLUP $\hat{\theta} _{i}$ is in the form of a composite estimate
[equation (\ref{4.4})], but with a tuning (shrinkage) coefficient $\gamma_{i} =
\sigma_{u}^{2}/(\sigma_{u}^{2} + \sigma_{Di}^{2})$, which is a function
of the ratio $\sigma_{u}^{2}/\sigma_{Di}^{2}$ of the variances of the
prediction errors of $\mathrm{x}'_{i}\beta$ and $\tilde{y}_{i}$,
respectively. The coefficient $\gamma_{i}$ defines optimally the
weights
assigned to the synthetic estimator $\mathrm{x}'_{i}\hat{\beta} _{\mathrm{GLS}}$
and $\tilde{y}_{i}$, unlike the case of design-based estimators where
the weight is assigned in a more ad hoc manner. See the discussion below
(\ref{4.4}). Note that the BLUP property does not require specifying the
distribution of the error terms beyond the first two moments, and
$\hat{\theta} _{i}$ is also the linear Bayes predictor in this case.
Under normality of the error terms and a diffuse uniform prior for
$\beta, \hat{\theta} _{i}$ is the Bayesian predictor (posterior mean) of
$\theta_{i}$. For a nonsampled area $k$, the BLUP is now\vspace*{1pt} obtained
optimally as $\mathrm{x}'_{k}\hat{\beta} _{\mathrm{GLS}}$.

In practice, the variance $\sigma_{u}^{2}$ is seldom known and is
replaced in $\gamma_{i}$ and $\hat{\beta} _{\mathrm{GLS}}$ by a sample estimate,
yielding what is known as the empirical BLUP (EBLUP) under the
frequentist approach, or the empirical\break Bayes (EB) predictor when
assuming normality. The latter predictor is the posterior mean of
$\theta_{i}$, but with $\sigma_{u}^{2}$ replaced by a sample estimate
obtained from the marginal distribution of the direct estimators given
the variance. Alternatively, one may compute the Hierarchical Bayes (HB)
predictor by assuming prior distributions for $\beta$ and $\sigma_{u}^{2}$ and
computing the posterior distribution of $\theta_{i}$ given the available
data. The posterior distribution can be used for computation of the point predictor and a
credibility (confidence) interval.

\begin{Remark}\label{rem1}
 The synthetic estimator
$\mathrm{x}'_{i}\hat{\beta} _{\mathrm{GLS}}$, and hence the BLUP $\hat{\theta}
_{i}$ are unbiased predictors under the joint distribution of $y_{i}$
and $\theta_{i}$ in the sense that $E(\hat{\theta} _{i} - \theta_{i}) =
0$, but are biased when conditioning on $u_{i}$. The predictor
$\hat{\theta} _{i}$ is biased also under the randomization distribution.
Conditioning on $u_{i}$ amounts to assuming different fixed intercepts
in different areas and the unbiasedness of $\hat{\theta} _{i}$ under the
model is achieved by viewing the intercepts as random.
\end{Remark}

\begin{Remark}\label{rem2}
 It is often the case that the \textit{linking
model} is defined for a transformation of $\theta_{i}$. For example, \citet{FaHe79}
actually assume $\log(\theta_{i}) =
\mathrm{x}'_{i}\beta + u_{i}$ in (\ref{5.1}) and use the direct estimator
$\tilde{y}_{i} = \log(\bar{y}_{i})$, and then predict $\theta_{i}$ as
$\exp(\tilde{\theta} _{i})$, where $\tilde{\theta} _{i}$ is the BLUP
(EBLUP) of $\log(\theta_{i})$ under the model. However,
$\exp(\tilde{\theta} _{i})$ is not the BLUP of $\theta_{i} = \exp
[\log(\theta_{i})]$. On the other hand, the EB and HB approaches produce
optimal predictors of $\theta_{i}$, even if the linking model uses a
transformation of $\theta_{i}$, with or without the use of a similar
transformation for the direct estimator. In this respect, the latter two
approaches are more flexible and with wider applicability, but at the
expense of requiring further parametric assumptions.
\end{Remark}

\subsubsection{Nested error unit level model}

This model uses individual observations $y_{ij}$ such that $y_{i}$ is
now a vector, and $\mathrm{x}_{i}$ is generally a matrix. The use of
this model for SAE requires that the area means
$\mathrm{\bar{X}}_{i}=\sum_{j = 1}^{N_{i}}
\mathrm{x}_{ij}/N_{i}$ are known. The model, first proposed for SAE by
Battese, Harter and Fuller (\citeyear{BaHaFu88}) has the form
\begin{equation}\label{5.3}
y_{ij} = \mathrm{x}'_{ij}\beta + u_{i} + \varepsilon_{ij},
\end{equation}
where the $u_{i}$s (random effects) and the $\varepsilon_{ij}$s
(residual terms) are mutually independent with zero means and variances
$\sigma_{u}^{2}$ and $\sigma_{\varepsilon} ^{2}$, respectively. Under
the model, the true small area means are $\bar{Y}_{i} =
\bar{X}'_{i}\beta + u_{i} + \bar{\varepsilon} _{i}$, but since
$\bar{\varepsilon} _{i} = \sum_{j = 1}^{N_{i}} \varepsilon_{ij}/N_{i}
\cong 0$ for large $N_{i}$, the target means are often defined as
$\theta_{i} = \bar{X}'_{i}\beta + u_{i} = E(\bar{Y}_{i}|u_{i})$. For
known variances $(\sigma_{u}^{2},\sigma_{\varepsilon} ^{2})$, the BLUP
of $\theta_{i}$ is
\begin{eqnarray}\label{5.4}
\hat{\theta} _{i} &=& \gamma_{i}[\bar{y}_{i} +
(\bar{X}_{i} - \bar{\mathrm{x}}_{i})
'\hat{\beta} _{\mathrm{GLS}}]
\nonumber\\[-8pt]\\[-8pt]
&&{} + (1 -
\gamma_{i})\bar{X}'_{i}\hat{\beta} _{\mathrm{GLS}},\nonumber
\end{eqnarray}
where $\hat{\beta} _{\mathrm{GLS}}$ is the GLS of $\beta$ computed from all the
observations, $\mathrm{\bar{x}}_{i} = \sum_{j = 1}^{n_{i}}
\mathrm{x}_{ij}/n_{i}$ and $\gamma_{i} = \sigma_{u}^{2}/(\sigma_{u}^{2}
+ \sigma_{\varepsilon} ^{2}/n_{i})$. For area $k$ with no sample (but
known $\bar{X}_{k})$, the BLUP is $\hat{\theta} _{k}=\mathrm{
\bar{X}'}_{k}\hat{\beta} _{\mathrm{GLS}}$. See \citet{Ra03} for the BLUP of the
means $\bar{Y}_{i}$ in sampled areas.

The BLUP (\ref{5.4}) is also the Bayesian predictor (posterior mean) under
normality of the error terms and a diffuse uniform prior for $\beta$.
Replacing the variances $\sigma_{u}^{2}$ and $\sigma_{\varepsilon} ^{2}$
in $\gamma_{i}$ and $\hat{\beta} _{\mathrm{GLS}}$ by sample estimates yields the
corresponding EBLUP or EB predictors. Hierarchical Bayes (HB) predictors
are obtained by specifying prior distributions for $\beta$ and the two
variances and computing the posterior distribution of $\theta_{i}$ (or
$\bar{Y}_{i})$ given all the sample observations in all the areas.
Remark~\ref{rem1} applies to the BLUP (EBLUP) under this model as well.

\subsubsection{Mixed logistic model}

The previous two models assume continuous responses. Suppose now that
$y_{ij}$ is binary, taking the values 1 or 0, in which case the small
area quantities of interest are usually proportions or counts (say, the
proportion or total of unemployed persons in the area). The following
generalized linear mixed model (GLMM) considered originally by MacGibbon
and Tomberlin (\citeyear{MaTo89}) for SAE is in broad use for this kind of problems:
\begin{eqnarray}\label{5.5}
\hspace*{27pt}\operatorname{Pr}(y_{ij} = 1|p_{ij}) &=& p_{ij} ;\hspace*{-27pt}\nonumber\\[-8pt]\\[-8pt]
 \operatorname{logit}(p_{ij}) &=&
\mathrm{x}'_{ij}\beta + u_{i} ;\quad u_{i}\sim N(0,\sigma_{u}^{2}).\nonumber
\end{eqnarray}
The responses $y_{ij}$ are assumed to be conditionally independent,
given the random effects $u_{i}$, and likewise for the random effects.
The purpose is to predict the true area proportions $p_{i} = \sum_{j =
1}^{N_{i}} y_{ij}/N_{i}$. Let $\psi = (\beta,\sigma_{u}^{2})$ denote the
model parameters. For this model, there is no explicit expression for
the best predictor (BP) under a quadratic loss function, that is, for
$\hat{p}_{i}^{\mathrm{BP}} = E(p_{i}|y_{i},\mathrm{x}_{i};\psi)$, but as shown in
\citet{JiLa06b}, the BP can be computed (approximated)
numerically as the ratio of two one-dimensional integrals. Jiang and
Lahiri
review methods of estimating $\psi$, yielding the empirical BP (EBP)
$\hat{p}_{i}^{\mathrm{EBP}} = E(p_{i}|y_{i},\mathrm{x}_{i};\hat{\psi} )$, which
is also the EB predictor under the same assumptions. Application of the
full HB approach under this model consists of the following basic steps:
\begin{longlist}[(3)]
\item[(1)] specify prior distributions for $\sigma_{u}^{2}$ and $\beta $;

\item[(2)] generate observations from the posterior distributions of $\beta$, $\sigma_{u}^{2}$ and
$u_{1},\ldots,u_{m}$ by say, MCMC simulations, and draw a large number of
realizations $(\hat{\beta} ^{(r)},\sigma_{u}^{2(r)},\{
\hat{u}_{i}^{(r)}\} )$, $r = 1,\ldots,R$, $i = 1,\ldots,m$, and hence realizations
$y_{ik}^{(r)}\sim p_{ik}^{(r)} = \frac{\exp (x'_{ik}\beta ^{(r)} +
u_{i}^{(r)})}{1 + \exp (x'_{ik}\beta ^{(r)} + u_{i}^{(r)})}$ for $k
\notin s_{i}$;

\item[(3)] predict: $\hat{p}_{i} = (\sum_{j \in s_{i}} y_{ij} + \sum_{k \notin
s_{i}} \hat{y}_{ik})/N_{i}$; $\hat{y}_{ik} = \sum_{r = 1}^{R}
y_{ik}^{(r)}/R , k \notin s_{i}$.
\end{longlist}

Writing $\hat{p}_{i} = \frac{1}{R}\sum_{r = 1}^{R} (\sum_{j \in s_{i}}
y_{ij} + \sum_{k \notin s_{i}} {y}_{ik}^{(r)}) /N_{i}=
\frac{1}{R}\sum_{r = 1}^{R} \hat{p}_{i}^{(r)}$, the posterior variance
is approximated as $\hat{V}_{\mathrm{post}}(\hat{p}_{i}) = \frac{1}{R(R -
1)}\sum_{r = 1}^{R} (\hat{p}_{i}^{(r)} - \hat{p}_{i})^{2}$\vspace*{1pt}.

Ghosh et al. (\citeyear{GhNaStCa98}) discuss the use of HB SAE for GLMM, covering binary,
count, multi-category and spatial data. In particular, sufficient
conditions are developed for the joint posterior distribution of the
parameters of interest to be proper.

\section{New Developments in Model-Based SAE}\label{sec6}

\subsection{Estimation of Prediction MSE}\label{sec6.1}

As stated in the introduction, an important aspect of SAE is the
assessment of the accuracy of the predictors. This problem is solved
``automatically'' under the Bayesian paradigm, which produces
realizations of the posterior distribution of the target quantities.
However, estimation of the prediction MSE (PMSE) and the computation of
confidence intervals (C.I.) under the frequentist approach is
complicated because of the added variability induced by the estimation
of the model hyper-parameters. \citet{PrRa90} developed PMSE
estimators with bias of order $o(1/m)$, ($m$~is the number of sampled
areas), under the linear mixed models (\ref{5.1}) and (\ref{5.2}) for the case where
the random errors have a normal distribution, and the model variances
are estimated by the ANOVA method of moments. \citet{DaLa00}
extended the estimation of Prasad and Rao to the more general mixed linear
model,
\begin{equation}\label{6.1}
y_{i} = X_{i}\beta + Z_{i}u_{i} + e_{i},\quad i = 1, \ldots, m,
\end{equation}
where $X_{i}$ and $Z_{i}$ are fixed matrices of order $n_{i} \times
k$ and $n_{i} \times d$, respectively, and $u_{i}$ and $e_{i}$ are
independent normally distributed random effects and residual terms of
orders $d \times 1$ and $n_{i} \times 1$, respectively, $u_{i}\sim
N_{d}(0,Q_{i})$, $e_{i}\sim N_{n_{i}}(0,R_{i})$. The variance matrices are
known functions of variance components $\zeta =
(\zeta_{1},\ldots,\zeta_{L})$. The authors develop PMSE estimators with
bias of order $o(1/m)$ for the EBLUP obtained when estimating $\beta$
and $\zeta$ by MLE or REML. Das, Jiang and Rao (\citeyear{DaJiRa04}) extend the model of
\citet{DaLa00} by relaxing the assumption of independence of
the error terms between the areas and likewise develop an estimator for
the PMSE of the EBLUP when estimating the unknown model parameters by
MLE or REML, with bias of order $o(1/m)$. Datta, Rao and Smith (\citeyear{DaRaSm05}) show that for the area level model (\ref{5.1}), if
$\sigma_{u}^{2}$ is estimated by the method proposed by \citet{FaHe79}, it is required to add an extra term to the PMSE estimator to
achieve the desired order of bias of $o(1/m)$. See \citet{Da09} for an extensive review of methods of estimating the PMSE of the EBLUP and
EB under linear mixed models (LMM).

Estimation of the PMSE under the GLMM is more involved, and in what
follows, I review resampling procedures that can be used in such cases.
For convenience, I consider the mixed logistic model (\ref{5.5}), but the
procedures are applicable to other models belonging to this class. The
first procedure, proposed by Jiang, Lahiri and Wan (\citeyear{JiLaWa02}) uses the jackknife
method. Let $\lambda_{i} = E(\hat{p}_{i}^{\mathrm{EBP}} - p_{i})^{2}$ denote the
PMSE, where $p_{i} = \sum_{j = 1}^{N_{i}} y_{ij}/N_{i}$ is the true
proportion and $\hat{p}_{i}^{\mathrm{EBP}} =
E(p_{i}|y_{i},\mathrm{x}_{i};\hat{\psi} )$ is the EBP. The following
decomposition holds:
\begin{eqnarray}\label{6.2}
\hspace*{15pt}\lambda_{i} &=& E\bigl(\hat{p}_{i}^{(\mathrm{BP})} - p_{i}\bigr)^{2} +
E\bigl(\hat{p}_{i}^{(\mathrm{EBP})} - \hat{p}_{i}^{(\mathrm{BP})}\bigr)^{2}\hspace*{-15pt}\nonumber\\[-8pt]\\[-8pt]
& =& M_{1i} + M_{2i},\nonumber
\end{eqnarray}
where $M_{1i}$ is the PMSE of the BP (assumes known parameter values)
and $M_{2i}$ is the contribution to the PMSE from estimating the model
parameters, $\psi$. Denote by $\hat{\lambda} _{i}^{\mathrm{BP}}(\hat{\psi} )$ the
``\textit{naive}'' estimator of $M_{1i}$, obtained by setting
$\psi = \hat{\psi}$. Let $\hat{\lambda} _{i}^{\mathrm{BP}}(\hat{\psi} _{ - l})$
denote the naive estimator when estimating $\psi$ from all the areas
except for area $l$, and $\hat{p}_{i}^{\mathrm{EBP}}(\hat{\psi} _{ - l})$ denote
the corresponding EBP. The jackknife estimator of PMSE is
\begin{eqnarray}\label{6.3}
\qquad\hat{\lambda} _{i}^{\mathrm{JK}} &=& \hat{M}_{1i} + \hat{M}_{2i};\nonumber
\\
\hat{M}_{1i} &=& \hat{\lambda} _{i}^{\mathrm{BP}}(\hat{\psi} )\nonumber\\[-8pt]\\[-8pt]
&&{} -
\frac{m - 1}{m}\sum_{l = 1}^{m} [\hat{\lambda} _{i}^{\mathrm{BP}}(\hat{\psi} _{ -
l}) - \hat{\lambda} _{i}^{\mathrm{BP}}(\hat{\psi} )], \nonumber\\
\hat{M}_{2i} &=& \frac{m -
1}{m}\sum_{l = 1}^{m} [\hat{p}_{i}^{\mathrm{EBP}}(\hat{\psi} _{ - l}) -
\hat{p}_{i}^{\mathrm{EBP}}(\hat{\psi} )]^{2}.\nonumber
\end{eqnarray}
Under some regularity conditions, $E(\hat{\lambda} _{i}^{\mathrm{JK}}) -
\lambda_{i} = o(1/m)$, as desired.

The jackknife estimator estimates the unconditional PMSE over the joint
distribution of the random effects and the responses. \citet{LoRa09} proposed a modification of the jackknife, which is simpler and
estimates the conditional PMSE,\break $E[(\hat{p}_{i}^{(\mathrm{EBP})}
\!-\! p_{i})^{2}|y_{i}]$. Denoting $q_{i}(\psi,y_{i})\! =\!
\operatorname{Var}(p_{i}|y_{i};\psi)$, the modification consists of replacing
$\hat{M}_{1i}$ in (\ref{6.3}) by $\hat{M}_{1i,c} = q_{i}(\hat{\psi} ,y_{i}) -
\sum_{l \ne i}^{m} [q_{i}(\hat{\psi} _{ - l},y_{i}) - q_{i}(\hat{\psi}
,y_{i})]$. The modified estimator $\hat{\lambda} _{i,c}^{\mathrm{JK}} =
\hat{M}_{1i,c} + \hat{M}_{2i}$ has bias of order $o_{p}(1/m)$ in
estimating the conditional PMSE and bias of order $o(1/m)$ in estimating
the unconditional PMSE.

\citet{HaMa06} propose estimating the\break PMSE by use of
double-bootstrap. For model (\ref{5.5}), the procedure consists of the
following steps:

(1) Generate a new population from the model (\ref{5.5}) with parameters
$\hat{\psi}$ and compute the ``true'' area proportions for this
population. Compute the EBPs based on new sample data and newly
estimated parameters. The new population and sample use the same
covariates as the original population and sample. Repeat the process
independently $B_{1}$ times, with $B_{1}$ sufficiently large. Denote by
$p_{i,b_{1}}(\hat{\psi} )$ and $\hat{p}_{i,b_{1}}^{(\mathrm{EBP})}(\hat{\psi}
_{b_{1}})$ the ``true'' proportions and corresponding EBPs for
population and sample $b_{1}$, $ b_{1} = 1,\ldots,B_{1}$. Compute the
first-step bootstrap PMSE estimator,
\begin{equation}\label{6.4}
\hat{\lambda} _{i,1}^{\mathrm{BS}} = \frac{1}{B_{1}}\sum_{b_{1} =
1}^{B_{1}} \bigl[\hat{p}_{i,b_{1}}^{(\mathrm{EBP})}(\hat{\psi} _{b_{1}}) -
p_{i,b_{1}}(\hat{\psi} )\bigr]^{2}.
\end{equation}

(2) For each sample drawn in Step (1), repeat the computations of Step
(1) $B_{2}$ times with $B_{2}$ sufficiently large, yielding new
``true'' proportions\break $p_{i,b_{2}}(\hat{\psi} _{b_{1}})$ and EBPs
$\hat{p}_{i,b_{2}}^{(\mathrm{EBP})}(\hat{\psi} _{b_{2}})$, $b_{2} = 1,\ldots,B_{2}$.
Compute the sec\-ond-step bootstrap PMSE estimator,
\begin{eqnarray}\label{6.5}
\hspace*{20pt}\hat{\lambda} _{i,2}^{\mathrm{BS}} &=& \frac{1}{B_{1}}\sum_{b_{1}}^{B_{1}}
\frac{1}{B_{2}}\hspace*{-20pt}\nonumber\\[-8pt]\\[-8pt]
&&\hspace*{31pt}{}\cdot\sum_{b_{2} = 1}^{B_{2}}
\bigl[\hat{p}_{i,b_{2}}^{(\mathrm{EBP})}(\hat{\psi} _{b_{2}}) - p_{i,b_{2}}(\hat{\psi}
_{b_{1}})\bigr]^{2}.\nonumber
\end{eqnarray}
The double-bootstrap PMSE estimator is obtained by computing one of the
classical bias corrected estimators. For example,
\begin{eqnarray}\label{6.6}
\hat{\lambda} _{i}^{D - \mathrm{BS}} = \cases{
 \hat{\lambda}
_{i,1}^{\mathrm{BS}} + (\hat{\lambda} _{i,1}^{\mathrm{BS}} - \hat{\lambda} _{i,2}^{\mathrm{BS}}
),\vspace*{2pt}\cr
 \quad\mbox{if } \hat{\lambda} _{i,1}^{\mathrm{BS}} \ge\hat{\lambda}
 _{i,2}^{\mathrm{BS}},
 \vspace*{2pt}\cr
\hat{\lambda} _{i,1}^{\mathrm{BS}}\exp [(\hat{\lambda} _{i,1}^{\mathrm{BS}} -
\hat{\lambda} _{i,2}^{\mathrm{BS}})/\hat{\lambda} _{i,2}^{\mathrm{BS}}],\vspace*{2pt}\cr
 \quad\mbox{if }
\hat{\lambda} _{i,1}^{\mathrm{BS}} < \hat{\lambda}
_{i,2}^{\mathrm{BS}}.
}
\end{eqnarray}
Notice that whereas the first-step bootstrap estimator (\ref{6.4}) has bias of
order\vadjust{\goodbreak} $O(1/m)$, the double-bootstrap estimator has bias of order
$o(1/m)$ under some regularity conditions.

\citet{PfCo12} develop a general method of bias
correction, which models the error of a target estimator as a function
of the corresponding bootstrap estimator, and the original estimators
and bootstrap estimators of the parameters governing the model fitted to
the sample data. This is achieved by drawing at random a large number of
plausible parameters governing the model, generating a pseudo original
sample for each parameter and bootstrap samples for each pseudo sample,
and then searching by a cross validation procedure the best functional
relationship among a set of eligible bias correction functions that
includes the classical bootstrap bias corrections. The use of this
method produces estimators with bias of correct order and under certain
conditions it also permits estimating the MSE of the bias corrected
estimator. Application of the method for estimating the PMSE under the
model (\ref{5.5}) in an extensive simulation study outperforms the
double-bootstrap and jackknife procedures, with good performance in
estimating the MSE of the PMSE estimators.

\begin{Remark}\label{rem3}
 All the resampling methods considered above are in
fact model dependent since they require computing repeatedly the
empirical best predictors under the model.
\end{Remark}

Chambers, Chandra and Tzavidis (\citeyear{ChChTz11}) develop conditional bias-robust PMSE estimators
for the case where the small area estimators can be expressed as
weighted sums of sample values. The authors assume that for unit $j
\in U_{i}$, $y_{j} = \mathrm{x}'_{j}\beta_{i} + e_{j}$; $E(e_{j}) = 0$,
$\operatorname{Var}(e_{j}) = \sigma_{j}^{2}$, $j = 1,\ldots,n_{i}$, with $\beta_{i}$ taken as
a fixed vector of coefficients, and consider linear estimators of the
form $\hat{\theta} _{i} = \sum_{k \in s} w_{ik} y_{k}$ with fixed
weights $w_{ik}$. Thus, if $\theta_{i}$ defines the true area mean,
\begin{eqnarray}\label{6.7}
\mathrm{Bias}_{i} &=& E(\hat{\theta} _{i} - \theta_{i})\nonumber\\
& =&
\Biggl(\sum_{h = 1}^{m} \sum_{j \in s_{h}} w_{ij}\mathrm{x}_{j}' \beta_{h}\Biggr) -
\bar{X}_{i}\beta_{i}, \\
\operatorname{Var}_{i} &=& \operatorname{Var}(\hat{\theta} _{i} - \theta_{i})\nonumber\\
& =&
N_{i}^{ - 2}\Biggl(\sum_{h = 1}^{m} \sum_{j \in s_{h}}
a_{ij}^{2}\sigma_{j}^{2} + \sum_{j \in r_{i}} \sigma_{j}^{2} \Biggr),\nonumber
\end{eqnarray}
where $r_{i} = U_{i} - s_{i}$ and $a_{ij} = N_{i}w_{ij} - I(j \in
U_{i})$, with $I( \cdot)$ defining the indicator function. Assuming that
for $j \in U_{i}$, $\mu_{j} = E(y_{j}|\mathrm{x}_{j}) =
\mathrm{x}_{j}'\beta_{i}$ is estimated as $\hat{\mu} _{j} =
\mathrm{x}_{j}'\hat{\beta} _{i} = \sum_{k \in s} \phi_{kj}y_{k}$ and
$\sigma_{j}^{2} \equiv\sigma^{2}$, the bias\vadjust{\goodbreak} and variance in (\ref{6.7}) are
estimated as
\begin{eqnarray}\label{6.8}
\hat{\mathrm{Bias}}_{i} &=& \Biggl(\sum_{h = 1}^{m} \sum_{j \in s_{h}} w_{ij}
\hat{\mu} _{j}\Biggr) - N_{i}^{ - 1}\sum_{j \in U_{i}} \hat{\mu} _{j} ,\nonumber\\
\hat{\operatorname{Var}}_{i} &=& N_{i}^{ - 2}\sum_{j \in s} [a_{ij}^{2} + (N_{i} -
n_{i})n_{i}^{ - 1}]\\
&&\hspace*{35pt}{}\cdot\lambda_{j}^{ - 1} (y_{j} - \hat{\mu} _{j})^{2},\nonumber
\end{eqnarray}
where $\lambda_{j} = (1 - \phi_{jj})^{2} + \sum_{k \in s( - j)}
\phi_{kj}^{2}$, and $s( - j)$ defines the sample without unit $j$.

The authors apply the procedure for estimating the PMSE of the EBLUP and
the MBDE estimator (\ref{4.11}) under model (\ref{5.3}), and for estimating the
PMSE of the M-quantile estimator defined in Section~\ref{sec6.6}. For the first two applications the authors condition
on the model variance estimators so that the PMSE
estimators do not have bias of desired order even under correct model
specification. On the other hand, the estimators are shown empirically
to have smaller bias than the traditional PMSE estimators in the
presence of outlying observations, although with larger MSEs than the
traditional estimators in the case of small area sample sizes.

\subsection{Computation of Prediction Intervals}\label{sec6.2}

As in other statistical applications, very often analysts are interested
in prediction intervals for the unknown area characteristics.
Construction of prediction intervals under the Bayesian approach, known
as \textit{credibility intervals}, is straightforward via the posterior
distribution of the predictor. A ``natural'' prediction interval
under the frequentist approach with desired coverage rate $(1 - \alpha)$
is $\hat{\theta} _{i}^{( \cdot )} \pm z_{\alpha
/2}[\hat{\operatorname{Var}}(\hat{\theta} _{i}^{( \cdot )} - \theta_{i})]^{1/2}$, where
$\hat{\theta} _{i}^{( \cdot )}$ is the EB, EBP or EBLUP predictor, and
$\hat{\operatorname{Var}}(\hat{\theta} _{i}^{( \cdot )} - \theta_{i})$ is an
appropriate estimate of the prediction error variance. However, even
under asymptotic normality of the prediction error, the use of this
prediction interval has coverage error of order $O(1/m)$, which is not
sufficiently accurate. Recent work in SAE focuses therefore on reducing
the coverage error via parametric bootstrap.

\citet{HaMa06} consider the following general model: for a
suitable smooth function $f_{i}(\beta)$ of the covariates $\mathrm{X}_{i} =
(\mathrm{x}_{i1},\ldots,\mathrm{x}_{in_{i}})$ in area $i$ and a vector parameter $\beta$,
random variables $\Theta_{i} = f_{i}(\beta) + u_{i}$; $E(u_{i}) = 0$ are
drawn from a distribution $Q\{ f_{i}(\beta);\xi\}$. The outcome
observations $y_{ij}$ are drawn independently from a distribution $R\{
l(\Theta_{i});\eta_{i}\}$, where $l( \cdot)$ is a known link function,
and $\eta_{i}$ is either known or is the same for every area $i$. For
given covariates $X_{i0}$, sample size $n_{i0}$ and known parameters,
an
$\alpha$-level prediction interval for the corresponding
realization $\Theta_{i0}$ is
\begin{equation}\label{6.9}
\quad I_{\alpha} (\beta,\xi) = \bigl[q_{(1 - \alpha )/2}(\beta,\xi),q_{(1 +
\alpha )/2}(\beta,\xi)\bigr],
\end{equation}
where $q_{\alpha} (\beta,\xi)$ defines the $\alpha $-level quantile of
the distribution $Q\{ f_{i}(\beta);\xi\}$. A naive prediction interval
with\vspace*{1pt} estimated parameters is ${I}_{\alpha} (\hat{\beta} ,\hat{\xi}
)$, but this interval has coverage error of order $O(1/m)$, and it does
not use the area-specific outcome values. To reduce the error,
${I}_{\alpha} (\hat{\beta} ,\hat{\xi} )$ is calibrated on $\alpha$.
This is implemented by generating parametric bootstrap samples and
re-estimating $\beta$ and $\xi$ similarly to the first step of the
double-bootstrap procedure for PMSE estimation described in Section~\ref{sec6.1}.
Denote by $\hat{I}_{\alpha} ^{*} = I_{\alpha} (\hat{\beta}
^{*},\hat{\xi} ^{*})$ the bootstrap interval, and let $\hat{\alpha}$
denote the solution of the equation $\Pr(\theta_{i}^{*}
\in\hat{I}_{\hat{\alpha}} ^{*}) = \alpha$, where $\theta_{i}^{*}\sim Q\{
f_{i}(\hat{\beta} ),\hat{\xi} \}$. The bootstrap-calibrated prediction
interval with coverage error of order $O(m^{ - 2})$ is
${I}_{\hat{\alpha}} (\hat{\beta} ,\hat{\xi} )$.

Chatterjee, Lahiri and Li (\citeyear{ChLaLi08}) consider the general linear mixed model of Das, Jiang and Rao (\citeyear{DaJiRa04}),
mentioned in Section~\ref{sec6.1}: $Y = X\beta + Zu + e$, where $Y$
(of dimension $n$) signifies all the observations in all the
areas, $X_{n \times p}$ and $Z_{n \times q}$ are known matrices and $u$
and $e$ are independent vector normal errors of random effects and
residual terms with variance matrices $D(\psi)$ and $R(\psi)$, which are
functions of a $k$-vector parameter $\psi$. Note that this model and the
model of \citet{HaMa06} include as special cases the mixed linear
models defined by (\ref{5.1}) and (\ref{5.3}). The present model cannot handle
nonlinear mixed models [e.g., the GLMM (\ref{5.5})], which the Hall and
Maiti model can, but it does not require conditional independence of the
observations given the random effects, as under the Hall and Maiti
model.

The (parametric bootstrap) prediction interval of Chatterjee, Lahiri and Li
(\citeyear{ChLaLi08}) for a univariate linear combination $t = c'(X\beta + Zu)$ is
obtained by the following steps. First compute the conditional mean,
$\mu_{t}$ and variance $\sigma_{t}^{2}$ of $t|Y;\beta,\psi$. Next
generate new observations $y^{*} = X\hat{\beta} + Zu^{*} + e^{*}$, where
$u^{*}\sim N(0,D(\hat{\psi} ))$, $e^{*}\sim N(0,R(\hat{\psi} ))$. From
$y^{*}$, estimate $\hat{\beta} ^{*}$ and $\hat{\psi} ^{*}$ using the same
method as for $\hat{\beta}$ and $\hat{\psi}$, and compute $\hat{\mu}
_{t}^{*}$ and $\hat{\sigma} _{t}^{*}$ (same as $\mu_{t}$ and
$\sigma_{t}$, but with estimated parameters). Denote by $L_{n}^{*}$ the
bootstrap distribution of $(\hat{\sigma}^{*} _{t})^{ - 1}(t^{*} - \hat{\mu}
_{t}^{*})$, where $t^{*} = c'(X\hat{\beta} + Zu^{*})$, and let $d = (p + k)$
be the total number of unknown parameters. Then\vadjust{\goodbreak} as $d^{2}/n \to 0$ and
under some regularity conditions, if $q_{1}$, $q_{2}$ satisfy
$L_{n}^{*}(q_{2}) - L_{n}^{*}(q_{1}) = 1 - \alpha$,
\begin{eqnarray}\label{6.10}
&&\Pr(\hat{\mu} _{t} + q_{1}\hat{\sigma} _{t} \le t \le\hat{\mu}
_{t} + q_{2}\hat{\sigma} _{t})\nonumber\\[-8pt]\\[-8pt]
&&\quad = 1 - \alpha + O(d^{3}n^{ - 3/2}).\nonumber
\end{eqnarray}
Note that this theory allows $d$ to grow with $n$ and that the coverage
error is defined in terms of $n$ rather than~$m$, the number of sampled
areas, as under the \citet{HaMa06} approach. The total sample
size increases also as the sample sizes within the areas increase, and
not just by increasing $m$. By appropriate choice of $t$, the interval
(\ref{6.10}) is area specific.

\begin{Remark}\label{rem4}
 The article by Chatterjee, Lahiri and Li (\citeyear{ChLaLi08}) contains a
thorough review of many other prediction intervals proposed in the
literature.
\end{Remark}

\subsection{Benchmarking}\label{sec6.3}

Model-based SAE depends on models that can be hard to validate and if
the model is misspecified, the resulting predictors may perform poorly.
Benchmarking robustifies the inference by forcing the model-based
predictors to agree with a design-based estimator for an aggregate of
the areas for which the design-based estimator is reliable. Assuming
that the aggregation contains all the areas, the benchmarking equation
takes the general form,
\begin{equation}\label{6.11}
\sum_{i = 1}^{m} b_{i} \hat{\theta} _{i,\mathrm{model}} = \sum_{i =
1}^{m} b_{i} \hat{\theta} _{i,\mathrm{design}}.
\end{equation}
The coefficients $\{ b_{i}\}$ are fixed weights, assumed without loss of
generality to sum to 1 (e.g., relative area sizes). Constraint (\ref{6.9}) has
the further advantage of guaranteeing consistency of publication between
the model-based small area predictors and the design-based estimator for
the aggregated area, which is often required by statistical bureaus. For
example, the model-based predictors of total unemployment in counties
should add up to the design-based estimate of total unemployment in the
country, which is deemed accurate.

A benchmarking method in common use, often referred to as ratio or
pro-rata adjustment, is
\begin{eqnarray}\label{6.12}
\qquad\hat{\theta} _{i,\mathrm{Ratio}}^{\mathrm{bench}} &=& \Biggl(\sum_{j = 1}^{m} b_{j} \hat{\theta}
_{j,\mathrm{design}}\bigg/\sum_{j = 1}^{m} b_{j} \hat{\theta} _{j,\mathrm{model}}\Biggr)\nonumber\\[-8pt]\\[-8pt]
&&{}\cdot\hat{\theta} _{i,\mathrm{model}}.\nonumber
\end{eqnarray}
The use of this procedure, however, applies the same ratio correction
for all the areas,\vadjust{\goodbreak} irrespective of the precision of the model-based
predictors before benchmarking. As a result, the prorated predictor in a
given area is not consistent as the sample size in that area increases.
Additionally, estimation of the PMSE of the prorated predictors is not
straightforward. Consequently, other procedures have been proposed in
the literature.

Wang, Fuller and Qu (\citeyear{WaFuQu08}) derive benchmarked BLUP (BBLUP) under the area level
model (\ref{5.1}) as the predictors minimizing $\sum_{i = 1}^{m}
\varphi_{i}E(\theta_{i} - \hat{\theta} _{i}^{\mathrm{bench}}) ^{2}$ subject to
(\ref{6.11}), where the $\varphi_{i}$s are chosen positive weights. The BBLUP
is
\begin{eqnarray}\label{6.13}
\hspace*{26pt}\hat{\theta} _{i,\mathrm{BLUP}}^{\mathrm{bench}} &=& \hat{\theta} _{i,\mathrm{model}}^{\mathrm{BLUP}}\hspace*{-26pt}
\nonumber\\
&&{}+ \delta_{i}\sum_{j = 1}^{m} b_{j}(\theta_{j,\mathrm{design}} - \hat{\theta}
_{j,\mathrm{model}}^{\mathrm{BLUP}});\\
 \delta_{i} &=& \Biggl(\sum_{j = 1}^{m} \varphi_{j}^{ -
1}b_{j}^{2}\Biggr)^{ - 1}\varphi_{i}^{ - 1}b_{i}.\nonumber
\end{eqnarray}
When the variance $\sigma_{u}^{2}$ is unknown, it is replaced by its
estimator everywhere in (\ref{6.13}), yielding the empirical BBLUP. You and Rao
(\citeyear{YoRa02}) achieve ``automatic benchmarking'' for the unit level model (\ref{5.3})
by changing the estimator of $\beta$. Wang, Fuller and Qu (\citeyear{WaFuQu08}) consider a
similar procedure for the area level model.
Alternatively, the authors propose to augment the covariates $\mathrm{x}'_{i}$ to
$\mathrm{\tilde{x}'}_{i} = (\mathrm{x}'_{i},b_{i}\sigma_{Di}^{2})$.
(The variances $\sigma_{Di}^{2}$ are considered known under the area level model.) The use of
the augmented model yields a BLUP that likewise satisfies the benchmark constraint
(\ref{6.11}) and is more robust to omission of an important covariate from $\mathrm{x}_{i}$, provided
that the missing covariate is sufficiently correlated with the added covariate in $\tilde{\mathrm{x}}_{i}$.

\citet{PfTi06} add monthly benchmark constraints of the
form (\ref{6.11}) to the measurement (observation) equation of a time series
state-space model fitted jointly to the direct estimates in several
areas. Adding benchmark constraints to time series models is
particularly important since time series models are slow to adapt to
abrupt changes. The benchmarked predictor obtained under the augmented
time series model belongs to the family of predictors (\ref{6.13}) proposed by
Wang, Fuller and Qu (\citeyear{WaFuQu08}). By adding the constraints to the model equations,
the use of this approach permits estimating the variance of the
benchmarked estimators as part of the model fitting. The variance
accounts for the variances of the model error terms, the variances and
autocovariances of the sampling errors of the direct estimators and of
the benchmarks $\sum_{i = 1}^{m} b_{i} \hat{\theta} _{ti,\mathrm{direct}}$, $t =
1,2,\ldots,$ and the cross-covariances and autocovariances between the
sampling errors of the direct estimators and the benchmarks.

Datta et al. (\citeyear{DaGhStMa11}) develop Bayesian benchmarking by minimizing
\begin{eqnarray}\label{6.14}
&&\sum_{i = 1}^{m} \varphi_{i}E[(\theta_{i} - \hat{\theta}
_{i}^{\mathrm{bench}}) ^{2}|\hat{\theta} _{\mathrm{design}}]\quad \mbox{s.t.}\nonumber\\[-8pt]\\[-8pt]
&&\quad\sum_{i = 1}^{m} b_{i}
\hat{\theta} _{i}^{\mathrm{bench}} = \sum_{i = 1}^{m} b_{i} \hat{\theta}
_{i,\mathrm{design}},\nonumber
\end{eqnarray}
where $\hat{\theta} _{\mathrm{design}} = (\hat{\theta}
_{1,\mathrm{design}},\ldots,\hat{\theta} _{m,\mathrm{design}})'$. The solution of this
minimization problem is the same as (\ref{6.13}), but with $\hat{\theta}
_{k,\mathrm{model}}^{\mathrm{BLUP}}$ replaced everywhere by the posterior mean
$\hat{\theta} _{k,\mathrm{Bayes}}$. Denote the resulting predictors by
$\hat{\theta} _{i,\mathrm{Bayes}}^{\mathrm{bench},1}$. The use of these predictors has
the\break
drawback of ``over shrinkage'' in the sense that\break $\sum_{i = 1}^{m}
b_{i}(\hat{\theta} _{i,\mathrm{Bayes}}^{\mathrm{bench},1} - \bar{\hat{\theta}} _{b,
\mathrm{Bayes}}^{\mathrm{bench},1})^{2}< \sum_{i = 1}^{m} b_{i}E[(\theta_{i} -
\bar{\theta} _{b} )^{2}|\break\hat{\theta} _{\mathrm{design}}]$, where
$\bar{\hat{\theta}} _{b, \mathrm{Bayes}}^{\mathrm{bench},1} = \sum_{i = 1}^{m}
b_{i}\hat{\theta} _{i,\mathrm{Bayes}}^{\mathrm{bench},1}$ and $\bar{\theta} _{b} = \break
\sum_{i= 1}^{m} b_{i}\theta_{i}$. To deal with this problem, Datta et al.
(\citeyear{DaGhStMa11}) propose to consider instead the predictors $\hat{\theta}
_{i,\mathrm{Bayes}}^{\mathrm{bench},2}$, satisfying the constraints
\begin{eqnarray}\label{6.15}
&\displaystyle\sum_{i = 1}^{m} b_{i} \hat{\theta} _{i,\mathrm{Bayes}}^{\mathrm{bench},2}
=
\sum_{i = 1}^{m} b_{i} \hat{\theta} _{i,\mathrm{design}} ;&\nonumber\\[-8pt]\\[-8pt]
\quad&\displaystyle \sum_{i = 1}^{m} b_{i}
\Biggl(\hat{\theta} _{i,\mathrm{Bayes}}^{\mathrm{bench},2} - \sum_{i = 1}^{m} b_{i} \hat{\theta}
_{i,\mathrm{design}}\Biggr)^{2} =H,&\nonumber
\end{eqnarray}
where $H = \sum_{i = 1}^{m} b_{i} E[(\theta_{i} - \bar{\theta}
_{b})^{2}|\hat{\theta} _{\mathrm{design}}]$. The benchmarked predictors have now
the form
\begin{eqnarray}\label{6.16}
\hat{\theta} _{i,\mathrm{Bayes}}^{\mathrm{bench},2} &=& \sum_{i = 1}^{m} b_{i}
\hat{\theta} _{i,\mathrm{design}}\nonumber\\
&&{} + A_{\mathrm{CB}}(\hat{\theta} _{i,\mathrm{Bayes}} -
\bar{\hat{\theta}} _{\mathrm{Bayes}});\\
 A_{\mathrm{CB}}^{2} &=& H\Big/\sum_{i = 1}^{m}
b_{i}(\hat{\theta} _{i,\mathrm{Bayes}} - \bar{\hat{\theta}} _{\mathrm{Bayes}})^{2}.\nonumber
\end{eqnarray}
Notice that the development of the Bayesian benchmarked predictors is
general and not restricted to any particular model. The PMSE of the
benchmarked predictor can be estimated as $\hat{E}[(\hat{\theta}
_{i,\mathrm{Bayes}}^{\mathrm{bench},2} - \theta_{i})^{2}|\break\hat{\theta} _{\mathrm{design}}] =
\operatorname{Var}(\hat{\theta} _{i,\mathrm{Bayes}}|\hat{\theta} _{\mathrm{design}}) + (\hat{\theta}
_{i,\mathrm{Bayes}}^{\mathrm{bench},2} - \hat{\theta} _{i,\mathrm{Bayes}})^{2}$, noting that the
cross-product $E[(\hat{\theta} _{i,\mathrm{Bayes}}^{\mathrm{bench},2} - \break\hat{\theta}
_{i,\mathrm{Bayes}})(\hat{\theta} _{i,\mathrm{Bayes}} - \theta_{i})|\theta_{\mathrm{design}}] = 0$.

\citet{NaSa11} likewise consider Bayes\-ian benchmarking,
focusing on estimation of area proportions. Denoting by $c_{i}$ the
number of sample units in area $i$ having characteristic $C$, and by
$p_{i}$ the probability to have this characteristic, the authors assume
the beta-binomial hierarchical Bayesian model,
\begin{eqnarray}\label{6.17}
c_{i}|p_{i}&\sim& \operatorname{Binomial}(n_{i},p_{i});\nonumber\\
\hspace*{18pt}p_{i}|\mu,\tau&\sim& \operatorname{Beta}[\mu\tau,(1 - \mu)\tau ],\quad i = 1,\ldots,m,\hspace*{-18pt} \\
p(\mu,\tau) &=& (1 + \tau^{2})^{ - 1},\quad 0 < \mu < 1, \tau\ge 0.\nonumber
\end{eqnarray}
Let $\tilde{b}_{i} = n_{i}/n$. The benchmark constraint is defined as,
\begin{equation}\label{6.18}
\hspace*{15pt}\sum_{i = 1}^{m} \tilde{b}_{i} p_{i} = \theta ;\quad \theta\sim
\operatorname{Beta}[\mu_{0}\tau_{0},(1 - \mu_{0})\tau_{0}].\hspace*{-15pt}
\end{equation}
The authors derive the joint posterior distribution of the true
probabilities $\{ p_{i}, i = 1,\ldots,m\}$ under the unrestricted model
(\ref{6.17}), and the restricted model with (\ref{6.18}), and prove that it is
proper. Computational details are given. Different scenarios are
considered regarding the prior distribution of $\theta$. Under the first
scenario $\tau_{0} \to\infty$, implying that $\theta$ is a point mass at
$\mu_{0}$, assumed to be known.
Under a second scenario $\mu_{0}$ and $\tau_{0}$ are specified by the
analyst. In a third scenario $\mu_{0} = 0.5$, $\tau_{0} = 2$, implying
$\theta\sim \operatorname{Uniform}(0,1)$ (noninformative prior). Theoretical arguments
and empirical results show that the largest gain from using the
restricted model is under the first scenario where $\theta$ is
completely specified, followed by the second scenario with $\tau_{0}
\gg 2$. No gain in precision occurs under the third scenario with a
noninformative prior.

To complete this section, I mention a different frequentist benchmarking
procedure applied by Ugarte, Militino and Goicoa (\citeyear{UgMiGo09}). By this procedure, the small
area predictors in sampled and nonsampled areas under the unit level
model (\ref{5.3}) are benchmarked to a synthetic estimator for a region
composed of the areas as obtained under a linear regression model with
heterogeneous variances (but no random effects). The benchmarked
predictors minimize a weighted residual sum of squares (WRSS) under model (\ref{5.3})
among all the predictors satisfying the benchmark constraint. Notice
that the predictors minimizing the WRSS without the constraint are the optimal
predictors (\ref{5.4}). For known variances the benchmarked predictors are
linear, but in practice the variances are replaced by sample estimates.
The authors estimate the PMSE of the resulting empirical benchmarked
predictors by a single-step parametric bootstrap procedure.

\subsection{Accounting for Measurement Errors in the Covariates}\label{sec6.4}

\citet{YbLo08} consider the case where some or all the
covariates $\mathrm{x}_{i}$ in the area level model (\ref{5.1}) are unknown,
and one uses an estimator $\mathrm{\hat{x}}_{i}$ obtained from another
independent survey, with\linebreak[4] $\operatorname{MSE}_{D}(\mathrm{\hat{x}}_{i}) = C_{i}$ under
the sampling design. (For\break known covariates $\mathrm{x}_{ki}$, $C_{ki} =
0$.) Denoting the resulting predictor by $\hat{\theta} _{i}^{\mathrm{Err}}$, it
follows that for known $(\beta,\sigma_{u}^{2} )$,
\begin{equation}\label{6.19}
\hspace*{15pt}\operatorname{PMSE}(\hat{\theta} _{i}^{\mathrm{Err}}) = \operatorname{PMSE}(\hat{\theta} _{i}) + (1 -
\gamma_{i})^{2}\beta 'C_{i}\beta,\hspace*{-20pt}
\end{equation}
where $\operatorname{PMSE}(\hat{\theta} _{i})$ is the PMSE if one knew
$\mathrm{x}_{i}$. Thus, reporting $\operatorname{PMSE}(\hat{\theta} _{i})$ in this case
results in under-reporting the true PMSE. Moreover, if $\beta
'C_{i}\beta > \sigma_{u}^{2} + \sigma_{Di}^{2}$, $\operatorname{MSE}(\hat{\theta}
_{i}^{\mathrm{Err}}) > \sigma_{Di}^{2} = \operatorname{Var}_D(\tilde{y}_{i})$. The authors propose
therefore to use instead the predictor
\begin{eqnarray}\label{6.20}
\hspace*{22pt}\hat{\theta} _{i}^{\mathrm{Me}} &=& \tilde{\gamma} _{i}\tilde{y}_{i} + (1 -
\tilde{\gamma} _{i})\mathrm{\hat{x}'}_{i}\beta ;\hspace*{-22pt}\nonumber\\[-8pt]\\[-8pt]
 \tilde{\gamma} _{i} &=&
(\sigma_{u}^{2} + \beta 'C_{i}\beta)/(\sigma_{Di}^{2} + \sigma_{u}^{2} +
\beta 'C_{i}\beta).\nonumber
\end{eqnarray}
The predictor $\hat{\theta} _{i}^{\mathrm{Me}}$ minimizes the MSE of linear
combinations of $\tilde{y}_{i}$ and $\mathrm{\hat{x}'}_{i}\beta$.
Additionally, $E(\hat{\theta} _{i}^{\mathrm{Me}} - \theta_{i}) = (1 -
\tilde{\gamma} _{i})[E_{D}(\mathrm{\hat{x}}_{i}) -
\mathrm{x}_{i}]'\beta$, implying that the bias vanishes if
$\mathrm{\hat{x}}_{i}$ is unbiased for $\mathrm{x}_{i}$, and
$E(\hat{\theta} _{i}^{\mathrm{Me}} - \theta_{i})^{2} = \tilde{\gamma}
_{i}\sigma_{Di}^{2} \le\sigma_{Di}^{2}$. The authors develop estimators
for $\sigma_{u}^{2}$ and $\beta$, which are then
substituted in (\ref{6.20}) to obtain the corresponding empirical predictor.
The PMSE of the empirical predictor is estimated using the jackknife
procedure of Jiang, Lahiri and Wan (\citeyear{JiLaWa02}), described in Section~\ref{sec6.1}.

Ghosh, Sinha and Kim (\citeyear{GhSiKi06}) and Torabi, Datta and Rao (\citeyear{ToDaRa09}) study a different situation
of measurement errors. The authors assume that the true model is the
unit level model (\ref{5.3}) with a single covariate $x_{i}$ for all the units
in the same area, but $x_{i}$ is not observed, and instead, different
measurements $x_{ij}$ are obtained for different sampled units $j \in
s_{i}$. The sample consists therefore of the observations $\{
y_{ij},x_{ij}; i = 1,\ldots,m, j = 1,\ldots,n_{i}\}$. An example giving rise
to such a scenario is where $x_{i}$ defines the true level of air
pollution in the area and the $x_{ij}$'s represent pollution measures at
different sites in the area. It is assumed that $x_{ij} = x_{i} +
\eta_{ij}$; $x_{i}\sim N(\mu_{x},\sigma_{x}^{2})$, and
$(u_{i},\varepsilon_{ij},\eta_{ij})$ are independent normally
distributed random errors with zero means and variances
$\sigma_{u}^{2}$, $\sigma_{\varepsilon} ^{2}$ and $\sigma_{\eta} ^{2}$,
respectively. Since $x_{i}$ is random, this kind of measurement error is
called \textit{structural measurement error}. The difference between the
two articles is that Ghosh, Sinha and Kim (\citeyear{GhSiKi06})\vadjust{\goodbreak} only use the observations $\{
y_{ij}\}$ for predicting the true area means $\bar{Y}_{i}$, whereas
Torabi, Datta and Rao (\citeyear{ToDaRa09}) also use the sample observations $\{ x_{ij}\}$.

Assuming that all the model parameters are known, the posterior
distribution of the unobserved $y$-values in area $i$ is multivariate
normal, which under the approach of Torabi, Datta and Rao (\citeyear{ToDaRa09}) yields the
following Bayes predictor (also BLUP) for $\bar{Y}_{i}$:
\begin{eqnarray}\label{6.21}
\hat{\bar{Y}}_{i}^{B} &=& E(\bar{Y}_{i}|\{ y_{ij} ,
x_{ij}, j=1,\ldots,n_i\} ) \nonumber\\
&=& (1 - f_{i}A_{i})\bar{y}_{i} + f_{i}A_{i}(\beta_{0} +
\beta_{1}\mu_{x})\\
&&{} + f_{i}A_{i}\gamma_{xi}\beta_{1}(\bar{x}_{i}
- \mu_{x}),\nonumber
\end{eqnarray}
where $f_{i} = 1 - (n_{i}/N_{i})$, $\gamma_{xi} =
n_{i}\sigma_{x}^{2}(\sigma_{\eta} ^{2} + n_{i}\sigma_{x}^{2})^{ - 1}$
and $A_{i} = [n_{i}\beta_{1}^{2}\sigma_{x}^{2}\sigma_{\eta} ^{2} +
(n_{i}\sigma_{u}^{2} + \sigma_{\varepsilon} ^{2})v_{i}]^{ -
1}\sigma_{\varepsilon} ^{2}v_{i}$, with $v_{i} = (\sigma_{\eta} ^{2} +
n_{i}\sigma_{x}^{2})$. For large $N_{i}$ and small $(n_{i}/N_{i})$, the
PMSE of $\hat{\bar{Y}}_{i}^{B}$ is $E[(\hat{\bar{Y}}_{i}^{B} -
\bar{Y}_{i})^{2}|\{ y_{ij},x_{ij}\} ] =
A_{i}[\beta_{1}^{2}\sigma_{x}^{2} + \sigma_{u}^{2} -
n_{i}\beta_{1}^{2}\sigma_{x}^{4}v_{i}^{ - 1}]$. Estimating the model
parameters $\psi =
(\beta_{0},\beta_{1},\mu_{x},\sigma_{x}^{2},\sigma_{u}^{2},\sigma_{\eta} ^{2},\sigma_{\varepsilon}
^{2})$ by a method of moments (MOM) proposed by Ghosh, Sinha and Kim (\citeyear{GhSiKi06}) and
replacing them by their estimates yields the EB estimator, which is
shown to be asymptotically optimal in the sense that $m^{ - 1}\sum_{i =
1}^{m} E(\hat{\bar{Y}}_{i}^{\mathrm{EB}} - \hat{\bar{Y}}_{i}^{B})^{2} \to 0$ as
$m \to\infty$. The PMSE of the EB predictor is estimated by a weighted
jackknife procedure of \citet{ChLa02}.

The Bayes predictor of Ghosh, Sinha and Kim (\citeyear{GhSiKi06}) has a similar structure to
(\ref{6.21}), but without the correction term
$f_{i}A_{i}\gamma_{xi}\beta_{1}(\bar{x}_{i} - \mu_{x})$, and
with the shrinkage coefficient $A_{i}$ replaced by $\tilde{A}_{i} =
[n_{i}(\beta_{1}^{2}\sigma_{x}^{2} + \sigma_{u}^{2}) +
\sigma_{\varepsilon} ^{2}]^{ - 1}\sigma_{\varepsilon} ^{2}$ in the other
two terms. As noted above, the authors develop a MOM for estimating the
unknown model parameters to obtain the EB predictor and prove its
asymptotic optimality. They also develop an HB predictor with
appropriate priors for all the parameters. The HB predictor and
its PMSE are obtained by MCMC simulations.

\citet{GhSi07} consider the same unit level model as above with
sample observations $(\{ y_{ij}\} ,\break\{ x_{ij}\} )$, but assume that the
true covariate $x_{i}$ is a fixed unknown parameter, which is known as
\textit{functional measurement error}. The work by \citet{YbLo08} reviewed before also assumes a functional measurement error, but
considers the area level model. For known parameters and $x_{i}$, the
Bayes predictor takes now the simple form
\begin{eqnarray}\label{6.22}
\quad\hat{\bar{Y}}_{i}^{B} &=& E(\bar{Y}_{i}|\{ y_{ij},j=1,\ldots,n_i\} )\nonumber\\
& =&
(1 - f_{i}B_{i})\bar{y}_{i} + f_{i}B_{i}(\beta_{0} +
\beta_{1}x_{i});\\
 B_{i} &=& (n_{i}\sigma_{u}^{2} +
\sigma_{\varepsilon} ^{2})^{ - 1}\sigma_{\varepsilon} ^{2}.\nonumber
\end{eqnarray}
A pseudo-Bayes predictor (PB) is obtained by substituting the sample
mean $\bar{x}_{i}$ for $x_{i}$ in (\ref{6.22}). A~pseu\-do-empirical Bayes
predictor (PEB) is obtained by estimating all the other unknown model
parameters by the MOM developed in Ghosh, Sinha and Kim (\citeyear{GhSiKi06}). The authors show
the asymptotic optimality of the PEB, $m^{ - 1}\sum_{i = 1}^{m}
E(\bar{Y}_{i}^{\mathrm{PEB}} - \bar{Y}_{i}^{\mathrm{PB}})^{2} \to 0$ as $m \to\infty$.

Datta, Rao and Torabi (\citeyear{DaRaTo10}) propose to replace the estimator $\bar{x}_{i}$ of
$x_{i}$ by its maximum likelihood estimator (MLE) under the model. The
corresponding PB of $\bar{Y}_{i}$ (assuming that the other model parameters are known) is the same as the PB of \citet{GhSi07}, but with
$B_{i}$ replaced by $\tilde{B}_{i} = (n_{i}\sigma_{u}^{2} +
\sigma_{\varepsilon} ^{2} + \beta_{1}^{2}\sigma_{\eta} ^{2})^{ -
1}\sigma_{\varepsilon} ^{2}$. A~PEB predictor is obtained by replacing
the model parameters by the MOM estimators developed in Ghosh, Sinha and Kim
(\citeyear{GhSiKi06}), and it is shown to be asymptotically optimal under the same
optimality criterion as before. The PMSE of the PEB is estimated by the
jackknife procedures of Jiang, Lahiri and Wan (\citeyear{JiLaWa02}) described in Section~\ref{sec6.1} and
the weighted jackknife procedure of \citet{ChLa02}. The authors
report the results of a simulation study showing that their PEB
predictor outperforms the PEB of \citet{GhSi07} in terms of
PMSE. A modification to the predictor of \citet{YbLo08} is also
proposed.

\subsection{Treatment of Outliers}\label{sec6.5}

\citet{BeHu06} consider the area level mod\-el~(\ref{5.1}) from a
Bayesian perspective, but assume that the random effect or the sampling
error (but not both) have a nonstandardized Student's $t_{(k)}$ distribution. The $t$
distribution is often used in statistical modeling to account for
possible outliers because of its long tails. One of the models
considered by the authors is
\begin{eqnarray}\label{6.23}
u_{i}|\delta_{i},\sigma_{u}^{2}&\sim&
N(0,\delta_{i}\sigma_{u}^{2});\nonumber\\
 \delta_{i}^{ - 1}&\sim& \operatorname{Gamma} [k/2,(k -
2)/2],\\
 e_{i}&\sim& N(0,\sigma_{Di}^{2}),\nonumber
\end{eqnarray}
which implies $E(\delta_{i}) = 1$ and $u_{i}|\sigma_{u}^{2}\sim
t_{(k)}(0,\sigma_{u}^{2}(k - 2)/k)$. The coefficient $\delta_{i}$ is
distributed around 1, inflating or deflating the variance of $u_{i} =
\theta_{i} - \mathrm{x}'_{i}\beta$. A~large value $\delta_{i}$ signals
the existence of an outlying area mean $\theta_{i}$. The degrees of
freedom parameter, $k$, is taken as known. Setting $k = \infty$ is
equivalent to assuming the model (\ref{5.1}). The authors consider several
possible (small) values for $k$ in their application, but the choice of
an appropriate value depends on data exploration. Alternatively, the
authors assume model (\ref{6.23}) for the sampling error $e_{i}$ (with
$\sigma_{Di}^{2}$ instead of $\sigma_{u}^{2}$), in which case it is
assumed that $u_{i}\sim N(0,\sigma_{u}^{2})$. The effect of assuming the
model for the random effects is to push the small area predictor (the
posterior mean) toward the direct estimator, whereas the effect of
assuming the model for the sampling errors is to push the predictor
toward the synthetic part. The use of either model is shown empirically
to perform well in identifying outlying areas, but at present it is not
clear how to choose between the two models. \citet{HuBe06} extend
the approach to a bivariate area level model where two direct estimates
are available for every area, with uncorrelated sampling errors but
correlated random effects. This model handles a situation where
estimates are obtained from two different surveys.

Ghosh, Maiti and Roy (\citeyear{GhMaRo08}) likewise consider model (\ref{5.1}) and follow the EB
approach. The starting point in this study is that an outlying direct
estimate may arise either from a large sampling error or from an
outlying random effect. The authors propose therefore to replace the EB
predictor obtained from (\ref{5.2}) by the robust EB predictor,
\begin{eqnarray}\label{6.24}
\qquad\hspace*{7pt}\hat{\theta} _{i}^{\mathrm{Rob}} &=& \tilde{y}_{i} - (1 - \hat{\gamma}
_{i})\hat{V}_{i}\Psi_{G} [(\tilde{y}_{i} - \mathrm{x}'_{i}\hat{\beta}
_{\hat{\mathrm{GLS}}})\hat{V}_{i}^{ - 1}] ;
\nonumber
\\[-8pt]
\\[-8pt]
\nonumber
 \hat{V}_{i}^{2} &=&
\hat{\operatorname{Var}}(\tilde{y}_{i} - \mathrm{x}'_{i}\hat{\beta} _{\mathrm{GLS}}),
\end{eqnarray}
where $\hat{\beta} _{\hat{\mathrm{GLS}}}$ is the empirical GLS under the model
with estimated variance $\hat{\sigma} _{u}^{2}$, and $\Psi_{G}$ is the
Huber influence function $\Psi_{G}(t) = \operatorname{sign}(t)\min(G,|t|)$ for some
value $G > 0$. Thus, for large positive standardized residuals
$(\tilde{y}_{i} - \mathrm{x}_{i}'\hat{\beta} _{\hat{\mathrm{GLS}}})\hat{V}_{i}^{
- 1}$, the EB $\hat{\theta}^{\mathrm{EB}} _{i} = \tilde{y}_{i} - (1 - \hat{\gamma}
_{i})\hat{V}_{i}(\tilde{y}_{i} - \mathrm{x}'_{i}\hat{\beta}
_{\mathrm{GLS}})\hat{V}_{i}^{ - 1}$ under the model is replaced by $\hat{\theta}
_{i}^{\mathrm{Rob}} = \tilde{y}_{i} - (1 - \hat{\gamma} _{i})\hat{V}_{i}G$, and
similarly for large negative standardized residuals, whereas in other
cases the ordinary~EB, $\hat{\theta}^{\mathrm{EB}} _{i}$, is unchanged. The value $G$
may change from one area to the other, and it is chosen adaptively in
such a way that the excess Bayes risk under model (\ref{5.1}) from using the
predictor (\ref{6.24}) is bounded by some percentage point. Alternatively, $G$
may be set to some constant $1 \le G_{0} \le 2$, as is often found in
the robustness literature. The authors derive the PMSE of $\hat{\theta}
_{i}^{\mathrm{Rob}}$ under the model (\ref{5.1}) for the case where $\sigma_{u}^{2}$ is
estimated by MLE with bias of order $o(1/m)$, and develop an estimator
for the PMSE that is correct up to the order $O_{p}(1/m)$.

Under the approach of Ghosh, Maiti and Roy\break (\citeyear{GhMaRo08}), the EB predictor is replaced
by the robust predictor (\ref{6.24}), but the estimation of the unknown model
parameters and the development of the PMSE and its estimator are under
the original model (\ref{5.1}), without accounting for possible outliers. Sinha
and Rao (\citeyear{SiRa09}) propose to robustify also the estimation of the model
parameters. The authors consider the mixed linear model (\ref{6.1}), which
when written compactly for all the observations $y =
(y'_{1},\ldots,y'_{m})'$, has the form
\begin{eqnarray}\label{6.25}
y &=& X\beta + Zu + e,\nonumber\\
 E(u) &=& 0,\quad E(uu') = Q;\\
E(e) &=& 0,\quad E(ee') = R,\nonumber
\end{eqnarray}
where $u$ is the vector of random effects, and $e$ is the vector of
residuals or sampling errors. The matrices $Q$ and $R$ are block
diagonal with elements that are functions of a vector parameter $\zeta =
(\zeta_{1},\ldots,\zeta_{L})$ of variance components such that $V(y) = V =
ZQZ' + R = V(\zeta)$. The target is to predict the linear combination
$\tau = l'\beta + h'u$ by $\hat{\tau} = l'\hat{\beta} + h'\hat{u}$.
Under the model, the MLE of $\beta$ and $\zeta$ are obtained by solving
the normal equations $X'V^{ - 1}(y - X\beta) = 0$; $(y - X\beta)'V^{ -
1}\frac{\partial V}{\partial \zeta _{l}}V^{ - 1}(y - X\beta) - \operatorname{tr}(V^{ -
1}\frac{\partial V}{\partial \zeta _{l}}) = 0$, $l = 1,\ldots,\break L$. To account
for possible outliers, the authors propose solving instead
\[
X'V^{ - 1}U^{1/2}\Psi_{G}(r) = 0 ;
\]
\begin{eqnarray}\label{6.26}
\quad&& \Psi '_{G}(r)U^{1/2}V^{ -
1}\frac{\partial V}{\partial \zeta _{l}}V^{ - 1}U^{1/2}\Psi_{G}(r)\\
&&\quad{} -
\operatorname{tr}\biggl(V^{ - 1}\frac{\partial V}{\partial \zeta _{l}}c\mathrm{I}_{n}\biggr) = 0,\quad l= 1,\ldots,L,\nonumber
\end{eqnarray}
where $r = U^{ - 1/2}(y - X\beta)$, $U = \operatorname{Diag}[V]$, $\Psi_{G}(r) =
[\Psi_{G}(r_{1}),\Psi_{G}(r_{2}),\ldots]'$ with $\Psi_{G}(r_{k})$ defining
the Huber influence function, $\mathrm{I}_{n}$ is the identity matrix of
order $n$ and $c = E[\Psi_{G}^{2}(r_{k})]$ [$r_{k}\sim N(0,1)$]. Notice
that since $Q$ and $R$ are block diagonal, the normal equations and the
robust estimating equations can be written as sums over the $m$ areas.

Denote by $\hat{\beta} _{\mathrm{Rob}}$, $\hat{\zeta} _{\mathrm{Rob}}$ the solutions of
(\ref{6.26}). The random effects are predicted by solving
\begin{eqnarray}\label{6.27}
&&Z'\hat{R}^{ - 1/2}\Psi_{G}[\hat{R}^{ - 1/2}(y - X\hat{\beta}
_{\mathrm{Rob}} - Zu)]\nonumber\\[-8pt]\\[-8pt]
&&\quad{} - \hat{Q}^{ - 1/2}\Psi_{G}(\hat{Q}^{ - 1/2}u) = 0,\nonumber
\end{eqnarray}
where $\hat{R} = R(\hat{\zeta} _{\mathrm{Rob}})$,  $\hat{Q} = Q(\hat{\zeta} _{\mathrm{Rob}})$.
Sinha and Rao (\citeyear{SiRa09}) estimate the PMSE of the robust small area
predictors by application\vadjust{\goodbreak} of the first step of the double-bootstrap procedure
of \citet{HaMa06} (equation~6.4). The parameter estimates and the
predictors of the random effects needed for the application of the
bootstrap procedure are computed by the robust estimating equations
(\ref{6.26})--(\ref{6.27}), but the generation of the bootstrap samples is under the
original model with no outliers. The estimation of the PMSE can possibly
be improved by generating some outlying observations, thus reflecting
more closely the properties of the original sample.

\subsection{Different Models and Estimators for Further
Robustification}\label{sec6.6}

In this section I review four different approaches proposed in the
literature for further robustification of the inference by relaxing some
of the model assumptions or using different estimators. All four studies
focus on the commonly used area-level and/or unit-level models defined
by (\ref{5.1}) and (\ref{5.3}), respectively.

\subsubsection*{M-quantile estimation}
 Classical model-based SAE methods mo\-del the
expectations $E(y_{i}|\mathrm{x}_{i},u_{i})$ and\break $E(u_{i})$. \citet{ChTz06} and Tzavidis, Marchetti and Chambers
 (\citeyear{TzMaCh10}) propose modeling instead
the quantiles of the distribution $f(y_{i}|\mathrm{x}_{i})$, where for
now $y_{i}$ is a scalar. Assuming a linear model for the quantiles, this
leads to a family of models indexed by the coefficient $q \in (0,1)$; $q =
\Pr [y_{i} \le \mathrm{x}'_{i}\beta_{q}]$. In quantile regression the vector
$\beta_{q}$ is estimated by minimizing
\begin{eqnarray}\label{6.28}
&&\mathop{\min}_{\beta _{q}}\sum_{i = 1}^{n} \{ |y_{i} -
\mathrm{x}'_{i} \beta_{q}|\nonumber\\
&&\hspace*{34pt}{}\cdot[(1 - q){I}(y_{i} -
\mathrm{x}'_{i}\beta_{q} \le 0)\\
&&\hspace*{60pt}{} + q{I}(y_{i} -
\mathrm{x}'_{i}\beta_{q} > 0)]\}.\nonumber
\end{eqnarray}
\textit{M-quantile} regression uses influence functions for estimating
$\beta_{q}$ by solving the equations
\begin{eqnarray}\label{6.29}
&&\sum_{i = 1}^{n} \Psi_{q}(r_{iq})\mathrm{x}_{i} = 0;\nonumber\\
&&\quad r_{iq} =
(y_{i} - \mathrm{x}'_{i}\beta_{q}),\nonumber\\[-8pt]\\[-8pt]
\qquad &&\Psi_{q}(r_{iq}) = 2\Psi(s^{ -
1}r_{iq})[(1 - q){I}(r_{iq} \le 0)\nonumber\\
&&\hspace*{118pt}{} + q{I}(r_{iq} > 0)],\nonumber
\end{eqnarray}
where $s$ is a robust estimate of scale, and $\Psi$ is an\vspace*{1pt} appropriate
influence function. The (unique) solution $\hat{\beta} _{q}$ of (\ref{6.29})
is obtained by an iterative reweighted least square algorithm. Note that
each sample value $(y_{i},\mathrm{x}_{i})$ lies on one and only one of\vadjust{\goodbreak}
the quantiles $m_{q}(\mathrm{x}_{i}) = \mathrm{x}'_{i}\beta_{q}$ (which
follows from the fact that the quantiles are continuous in $q$).

How is the M-quantile theory used for SAE? Suppose that the sample
consists of unit level observations $\{ y_{ij},\mathrm{x}_{ij}; i =
1,\ldots,m, j = 1,\ldots,n_{i}\}$. Identify for unit $(i,j)$ the value
$q_{ij}$ such that $\mathrm{x}'_{ij}\hat{\beta} _{q_{ij}} = y_{ij}$. A
predictor of the mean $\theta_{i}$ in area $i$ is obtained by averaging
the quantiles $q_{ij}$ over the sampled units $j \in s_{i}$ and
computing
\begin{eqnarray}\label{6.30}
\hat{\theta} _{i}^{M} &=& N_{i}^{ - 1}\biggl(\sum_{j \in s_{i}} y_{ij} +
\sum_{k \notin s_{i}} \mathrm{x}'_{ik}\hat{\beta} _{\bar{q}_{i}} \biggr);\nonumber\\[-8pt]\\[-8pt]
\bar{q}_{i} &=& \sum_{j = 1}^{n_{i}} q_{ij}/n_{i}.\nonumber
\end{eqnarray}
Alternatively, one can average the vector coefficients
$\beta_{q_{ij}}$and replace $\hat{\beta} _{\bar{q}_{i}}$ in (\ref{6.30}) by the
mean $\bar{\hat{\beta}} _{i} =\break \sum_{j = 1}^{n_{i}} \hat{\beta}
_{q_{ij}}/n_{i}$. The vectors $\hat{\beta} _{\bar{q}_{i}}$ or
$\bar{\hat{\beta}} _{i}$ account for differences between the areas,
similarly to the random effects under the unit level model (\ref{5.3}).

The use of this approach is not restricted to the estimation of means,
although it does assume continuous $y$-values. For example, the
distribution function in area $i$ can be estimated as $\hat{F}_{i}(t)
=\break
N_{i}^{ - 1}[\sum_{j \in s_{i}} {I}(y_{ij} \le t) + \sum_{k
\notin s_{i}} {I}(\mathrm{x}'_{ik}\bar{\hat{\beta}} _{i} \le
t)]$. \citet{ChTz06} develop unconditional and area
specific estimators for the variance of the M-quantile estimators (\ref{6.30})
assuming $\hat{\beta} _{\bar{q}_{i}}$ (or $\bar{\hat{\beta}} _{i})$ is
fixed, and estimators for the bias under the linear model $E(y_{ij}|x_{ij}) =
x'_{ij}\beta_{i}$.

The M-quantile approach does not assume a parametric model, although it
assumes that the quantiles are linear in the covariates in the theory
outlined above. Clearly, if the unit level model (\ref{5.3}) holds, the use
of the model is more efficient, but the authors illustrate that the
M-quantile estimators can be more robust to model misspecification.
Notice in this regard that the approach is not restricted to a specific
definition of the small areas. It accounts also for possible outliers by
choosing an appropriate influence function in the estimating equation
(\ref{6.29}). On the other hand, there seems to be no obvious way of how to
predict the means or other target quantities for nonsampled areas. A
possible simple solution would be to set $q = 0.5$ for such areas or
weight the $q$-values of neighboring sampled areas, but it raises the
question of how to estimate the corresponding PMSE, unless under a
model.

\subsubsection*{Use of penalized spline regression}
 Another way of
robustifying the inference is by use of\vadjust{\goodbreak} penalized spline (P-spline)
regression. The idea here is to avoid assuming a specific functional
form for the expectation of the response variable. Suppose that there is
a single covariate $x$. The P-spline model assumes $y = m_{0}(x) +
\varepsilon$, $E(\varepsilon) = 0$, $\operatorname{Var}(\varepsilon) = \sigma_{\varepsilon}
^{2}$. The mean $m_{0}(x)$ is taken as unknown and approximated
as
\begin{eqnarray}\label{6.31}
m(x;\beta,\gamma) &=& \beta_{0} + \beta_{1}x
+ \cdots + \beta_{p}x^{p}\nonumber\\
&&{} + \sum_{k = 1}^{K} \gamma_{k}(x - K_{k})_{ +}
^{p};\\
 (x - K_{k})_{ +} ^{p} &=& \max [0,(x - K_{k})^{p}],\nonumber
\end{eqnarray}
where $p$ is the degree of the spline, and $K_{1} < \cdots < K_{K}$ are
fixed knots. For large $K$ and good spread of the knots over the range
of $x$, spline (\ref{6.31}) approximates well most smooth functions. It uses
the basis $[1,x,\ldots,x^{p},(x - K_{1})_{ +} ^{p},\ldots,(x - K_{K})_{ +}
^{p}]$ to approximate the mean, but other bases can be considered,
particularly when there are more covariates.

Opsomer et al. (\citeyear{OpClRaKaBr08}) use P-spline regression for SAE by treating the
$\gamma $-coefficients in (\ref{6.31}) as additional random effects. Suppose
that the data consist of unit-level observations, $\{
y_{ij},\mathrm{x}_{ij}; i = 1,\ldots,m, j = 1,\ldots,n_{i}\}$. For unit $j$ in
area $i$, the model considered is
\begin{eqnarray}\label{6.32}
y_{ij} &=& \beta_{0} + \beta_{1}x_{ij} + \cdots + \beta_{p}x_{ij}^{p}\nonumber\\[-8pt]\\[-8pt]
&&{}+ \sum_{k = 1}^{K} \gamma_{k}(x_{ij} - K_{k})_{ +} ^{p} + u_{i} +
\varepsilon_{ij},\nonumber
\end{eqnarray}
where the $u_{i}$s are the usual area random effects and
$\varepsilon_{ij}$s are the residuals. Let $u = (u_{1},\ldots,u_{m})'$,
$\gamma = (\gamma_{1},\ldots,\gamma_{K})'$. Defining $d_{ij} = 1$ (0) if unit
$j$ is (is not) in area $i$ and denoting $d_{j} = (d_{1j},\ldots,d_{mj})'$
and $D = [d_{1},\ldots,d_{n}]'$, the model holding for the vector $y$ of
all the response values can be written compactly as
\begin{eqnarray}\label{6.33}
y &=& X\beta + Z\gamma + Du + \varepsilon ;\nonumber\\
\gamma&\sim&(0,\sigma_{\gamma} ^{2}\mathrm{I}_{k}),\\
u&\sim&(0,\sigma_{u}^{2}\mathrm{I}_{m}),\quad
\varepsilon\sim(0,\sigma_{\varepsilon} ^{2}\mathrm{I}_{n}),\nonumber
\end{eqnarray}
where $X = [x_{1}^{(p)},\ldots,x_{n}^{(p)}]'$ with $x_{l}^{(p)} =
(1,x_{l},\ldots,\allowbreak x_{l}^{p})'$, and $Z = [z_{1},\ldots,z_{n}]'$ with $z_{l} =
[(x_{l} - K_{1})_{ +} ^{p},\ldots,\allowbreak (x_{l} - K_{K})_{ +} ^{p})]'$. The model
(\ref{6.33}) looks similar to (\ref{6.25}) but the responses $y_{ij}$ are not
independent between the areas because of the common random effects
$\gamma$. Nonetheless, the BLUP and EBLUP of $(\beta,u,\gamma)$ can be
obtained using standard results; see the article for the appropriate
expressions.\vadjust{\goodbreak} The small area EBLUP are obtained as
\begin{eqnarray}\label{6.34}
\hspace*{25pt}\hat{\theta} _{i,\mathrm{EBLUP}}^{\mathrm{P\mbox{-}spline}} &=& \hat{\beta}
'\bar{X}_{i}^{(p)} + \hat{\gamma} '\bar{Z}_{i} + \hat{u}_{i};\hspace*{-25pt}\nonumber\\[-8pt]\\[-8pt]
\bar{X}_{i}^{(p)} &=& \sum_{l \in U_{i}} x_{l}^{(p)} /N_{i},\quad \bar{Z}_{i} =
\sum_{l \in U_{i}} z_{l}/N_{i}.\nonumber
\end{eqnarray}

The use of this approach requires that the covariates are known for
every element in the population. Opsomer et al. (\citeyear{OpClRaKaBr08}) derive the PMSE
of the EBLUP~(\ref{6.34}) correct to second order for the case where the
unknown variances are estimated by REML, and an estimator of the PMSE
with bias correct to the same order. The authors develop also a
nonparametric bootstrap algorithm for estimating the PMSE and for
testing the hypotheses $\sigma_{u}^{2} = 0$ and $\sigma_{\gamma} ^{2} =
0$ of no random effects. \citet{RaSiRo09} use a similar model to
(\ref{6.33}), but rather than computing the EBLUP under the model, the authors
propose predictors that are robust to outliers, similar (but not the
same) to the methodology developed by Sinha and Rao (\citeyear{SiRa09}) for the mixed
linear model described in Section~\ref{sec6.5}. Jiang, Nguyen and Rao (\citeyear{JiNgRa10}) show how to
select an appropriate spline model by use of the fence method described in Section~\ref{sec8}.

\subsubsection*{Use of empirical likelihood in Bayesian inference}

\citet{ChGh11} consider the use of
empirical likelihoods (EL) instead of fully parametric likelihoods as
another way of robustifying the inference. When combined with
appropriate proper priors, it defines a semiparametric Bayesian
approach, which can handle continuous and discrete outcomes in area- and
unit-level models, without specifying the distribution of the outcomes as
under the classical Bayesian approach. Denote by $\theta =
(\theta_{1},\ldots,\theta_{m})'$ and $y = (y_{1},\ldots,y_{m})'$ the area
parameters and the corresponding 
%
direct estimators, and by $\tau = (\tau_{1},\ldots,\break\tau_{m})$ the ``jumps'' defining the
cumulative distribution of $y_i$, so that $\sum_{i = 1}^{m} \tau_{i} = 1$. The EL is $L_{E}= \prod_{i = 1}^{m} {\tau} _{i}$
and for given moments
$E(y_{i}|\theta_{i}) = k(\theta_{i})$,\break $\operatorname{Var}(y_{i}|\theta_{i}) =
V(\theta_{i})$, the estimate $\hat{\tau} (\theta)$ is the solution of
the constrained maximization problem
\begin{eqnarray}\label{6.35}
&&\mathop{\max} _{\tau _{1},\ldots,\tau _{m}}\prod_{i = 1}^{m}
\tau_{i} ,\nonumber\\
&&\quad \mbox{s.t.}\quad \tau_{i} \ge 0,\quad \sum_{i = 1}^{m} \tau_{i} = 1 ,\nonumber\\[-8pt]\\[-8pt]
 &&\hspace*{36pt}\sum_{i= 1}^{m} \tau_{i} [y_{i} - k(\theta_{i})] = 0,\nonumber\\
 &&\hspace*{36pt}\sum_{i = 1}^{m}
\tau_{i}\biggl\{ \frac{[y_{i} - k(\theta _{i})]^{2}}{V(\theta _{i})} - 1\biggr\} =
0.\nonumber
\end{eqnarray}
Under the area model (\ref{5.1}) $k(\theta_{i}) = \theta_{i} =
\mathrm{x}'_{i}\beta + u_{i}$ and $V(\theta_{i}) = \sigma_{Di}^{2}$. The
authors assume proper priors for $(\beta,
u_{1},\ldots,u_{m},\sigma_{u}^{2})$ and hence for $\theta$, thus
guaranteeing that the posterior distribution $\pi(\theta |y)$ is also
proper. For given $\theta$ the constrained maximization problem (\ref{6.35})
is solved by standard methods (see the article), and by combining the EL
with the prior distributions, observations from the posterior
distribution $\pi(\theta |y)$ are obtained by MCMC simulations.

For the unit-level model (\ref{5.3}), $E(y_{ij}|\theta_{ij}) = k(\theta_{ij})
= \mathrm{x}'_{ij}\beta + u_{i}$ and $\operatorname{Var}(y_{ij}|\theta_{ij}) =
V(\theta_{ij}) = \sigma_{\varepsilon} ^{2}$. Denoting by $\tau_{ij}$ the
``jumps'' of the cumulative distribution in area $i$, the EL is defined in
this case as $L_{E} = \prod_{i = 1}^{m} \prod_{j = 1}^{n_{i}}
{\tau} _{ij} = \prod_{i = 1}^{m} {\tau} _{(i)}$,
and for given $\theta_{(i)} = \break(\theta_{i1},\ldots,\theta_{i,n_{i}})'$,
$\hat{\tau} _{(i)}(\theta) = [\hat{\tau} _{i1}(\theta),\ldots,\hat{\tau}
_{in_{i}}(\theta)]'$ is the solution of the area specific maximization
problem
\begin{eqnarray}\label{6.36}
&&\mathop{\max} _{\{ \tau _{ij}\}} \prod_{j = 1}^{n_{i}}
\tau_{ij} ,\nonumber\\
&&\quad \mbox{s.t.}\quad \tau_{ij} \ge 0,\quad \sum_{j = 1}^{n_{i}} \tau_{ij} = 1
,\nonumber\\[-8pt]\\[-8pt]
&&\hspace*{38pt}\sum_{j = 1}^{n_{i}} \tau_{ij} [y_{ij} - k(\theta_{ij})] = 0,\nonumber\\
&&\hspace*{38pt}\sum_{j =
1}^{n_{i}} \tau_{ij}\biggl\{ \frac{[y_{ij} - k(\theta _{ij})]^{2}}{V(\theta
_{ij})} - 1\biggr\} = 0.\nonumber
\end{eqnarray}
The authors applied the procedure for estimating state-wise median
income of four-person families in the USA, using the area-level
model. Comparisons with the census values for the same year reveal much
better predictions under the proposed approach compared to the direct
survey estimates and the HB predictors obtained under normality of the
direct estimates.

\subsubsection*{Best predictive SAE}
 In the three previous ap-\break proaches~reviewed
 in this section, the intended robustification is achieved by
relaxing some of the model assumptions. Jiang, Nguyen and Rao (\citeyear{JiNgRa11}) propose
instead to change the estimation of the fixed model parameters. The idea
is simple. In classical model-based SAE the EBLUP or EB predictors are
obtained by replacing the parameters in the expression of the BP by
their MLE or REML estimators. Noting that in SAE the actual target is
the prediction of the area means, and the estimation of model parameters
is just an intermediate step, the authors propose to estimate the fixed
parameters in such a way that the resulting predictors are optimal under
some loss function.

Consider the area-level model (\ref{5.1}) with normal errors, and suppose
first that $\sigma_{u}^{2}$ is known. Under the model, $E(y_{i}) =
\mathrm{x}'_{i}\beta$, but suppose that the model is misspecified and
$E(y_{i}) = \mu_{i}$, such that $\theta_{i} = \mu_{i} + u_{i}$, $i =
1,\ldots,m$. Let $\tilde{\theta} _{i}$ be a predictor of $\theta_{i}$, and
define the mean square prediction error to be $\operatorname{MSPE}(\tilde{\theta} ) = \sum_{i = 1}^{m}
E(\tilde{\theta} _{i} - \theta_{i})^{2}$, where the expectation is under
the \textit{correct model}. By (\ref{5.2}), the MSPE of the BP for given
$\beta$ is $\operatorname{MSPE}[\tilde{\theta} (\beta)]= E\{ \sum_{i = 1}^{m}
[\gamma_{i}y_{i} + (1 - \gamma_{i})\mathrm{x}'_{i}\beta -
\theta_{i}]^{2} \}$. The authors propose minimizing the expression
inside the expectation with respect to $\beta$, which is shown to be
equivalent to minimizing $\sum_{i = 1}^{m} [(1 -
\gamma_{i})^{2}\mathrm{(x}'_{i}\beta)^{2} - 2\sum_{i = 1}^{m} (1 -
\gamma_{i})^{2}\mathrm{x}'_{i}\beta y_{i} ]$, yielding the ``best
predictive estimator'' (BPE)
\begin{eqnarray}
\hspace*{15pt}\tilde{\beta} &=& \Biggl[\sum_{i = 1}^{m} (1 - \gamma_{i})^{2}
\mathrm{x}_{i}\mathrm{x}'_{i}\Biggr]^{ - 1}\sum_{i = 1}^{m} (1 -
\gamma_{i})^{2} \mathrm{x}_{i}y_{i}.\hspace*{-15pt}
\end{eqnarray}
Notice that unless $\operatorname{Var}_{D}(e_{i}) = \sigma_{Di}^{2} = \sigma_{D}^{2}$,
$\tilde{\beta}$ differs from the commonly used GLS estimator under the
model (\ref{5.1}); $\hat{\beta} _{\mathrm{GLS}} = [\sum_{i = 1}^{m} \gamma_{i}
\mathrm{x}_{i}\mathrm{x}'_{i}]^{ - 1}\sum_{i = 1}^{m} \gamma_{i}
\mathrm{x}_{i}y_{i}$. The ``observed best predictor'' (OBP) of
$\theta_{i}$ is obtained by replacing $\hat{\beta} _{\mathrm{GLS}}$ by
$\tilde{\beta}$ in the BP (\ref{5.2}) under the model~(\ref{5.1}).

The authors derive also the BPE of $\psi = (\beta ',\sigma_{u}^{2})'$
for the case where $\sigma_{u}^{2}$ is unknown, in which case the OBP is
obtained by replacing $\sigma_{u}^{2}$ and $\hat{\beta} _{\mathrm{GLS}}$ by the
BPE of $\psi$ in (\ref{5.2}). Another extension is for the unit level
model~(\ref{5.3}), with the true area means and MSPE defined as $\theta_{i}
= \bar{Y}_{i}$ and $\operatorname{MSPE}[\tilde{\theta} (\psi)] = \sum_{i = 1}^{m} E_{D}
[\tilde{\theta} _{i}(\psi) - \theta_{i}]^{2}$, respectively, where\vspace*{2pt} $\psi
= (\beta ',\allowbreak \sigma_{u}^{2},\sigma_{\varepsilon} ^{2})'$ and $E_{D}(
\cdot)$ is the design (randomization) expectation over all possible
sample selections (Section~\ref{sec4.1}). The reason for using the design
expectation in this case is that it is almost free of model assumptions.
Theoretical derivations and empirical studies using simulated data and a
real data set illustrate that the OBP can outperform very significantly
the EBLUP in terms of PMSE if the underlying model is misspecified. The
two predictors are shown to have similar PMSE under correct model
specification.

\subsection{Prediction of Ordered Area Means}\label{sec6.7}

\citet{MaRi10} consider the following (hard) problem:
predict the ordered area means $\theta_{(1)} \le\theta_{(2)} \le\cdots
\le\theta_{(m)}$ under the area-level model $\tilde{y}_{i} = \mu + u_{i}
+ e_{i} = \theta_{i} + e_{i}$ [special case of (5.1)], with
$u_{i}\stackrel{\mathrm{i.i.d.}}{\sim}H(0,\sigma_{u}^{2})$, $e_{i}\stackrel{\mathrm{i.i.d.}}{\sim}G(0,\sigma_{e}^{2})$; $H$ and $G$ are general distributions
with zero means and variances $\sigma_{u}^{2}$ and $\sigma_{e}^{2}$. To
illustrate the difference between the prediction of ordered and
unordered means, consider the prediction of $\theta_{(m)} = \max_{i}\{
\theta_{i}\}$. If $\hat{\theta} _{i}$ satisfies $E(\hat{\theta}
_{i}|\theta_{i}) = \theta_{i}, i = 1,\ldots,m$, then $E[\max_{i}\{
\hat{\theta} _{i}\} |\{ \theta_{j}\} ] \ge\theta_{(m)}$ so that the
largest estimator overestimates the true largest mean. On the other
hand, the Bayesian predictors $\theta_{i}^{*} = E[\theta_{i}|\{
\hat{\theta} _{j}\} ]$ satisfy $E[\max_{i}\{ \theta_{i}^{*}\} ] <
E(\theta_{(m)})$, an underestimation in expectation.

\citet{WrStCr03} considered the prediction of ordered
means from a Bayesian perspective, but their approach requires heavy
numerical calculations and is sensitive to the choice of priors.
\citet{MaRi10} compare three predictors of the ordered
means under the frequentist approach, using the loss function
$L(\tilde{\theta} _{(\cdot )},\theta_{(\cdot)}) = \sum_{i = 1}^{m}
(\tilde{\theta} _{(i)} - \theta_{(i)})^{2}$ and the Bayes risk
$E[L(\tilde{\theta} _{(\cdot)},\theta_{(\cdot)})]$. Let $\hat{\theta} _{i}$
define the direct estimator of $\theta_{i}$ and $\hat{\theta} _{(i)}$
the $i$th ordered direct estimator (statistic). The predictors compared are
\begin{eqnarray}\label{6.38}
\tilde{\theta} _{(i)}^{(1)} &=& \hat{\theta} _{(i)};\nonumber\\
\hspace*{15pt}\tilde{\theta} _{(i)}^{(2)}(\delta) &=& \delta\hat{\theta} _{(i)} + (1 -
\delta)\bar{\hat{\theta}} ,\quad \bar{\hat{\theta}} = \sum_{i = 1}^{m}
\hat{\theta} _{i} /m;\hspace*{-15pt}\\
 \tilde{\theta} _{(i)}^{(3)} &=&
E\bigl(\theta_{(i)}|\hat{\theta} \bigr),\quad \hat{\theta} = (\hat{\theta}
_{1},\ldots,\hat{\theta} _{m})'.\nonumber
\end{eqnarray}
The results below assume that $\sigma_{u}^{2}$ and $\sigma_{e}^{2}$ are
known and that $\mu$ is estimated by $\bar{\hat{\theta}}$.

Denote by $\tilde{\theta} ^{[k]}_{(\cdot)}$ the predictor of the ordered means
when using the predictors $\tilde{\theta} _{(i)}^{(k)}$, $k = 1,2,3$, and
let $\gamma = \sigma_{u}^{2}(\sigma_{u}^{2} + \sigma_{e}^{2})^{ - 1}$ be
the shrinkage coefficient when predicting the unordered means [equation
(\ref{5.2})]. The authors derive several theoretical comparisons. For example, if $\gamma\le(m - 1)^{2}/(m +
1)^{2}$, then
\begin{eqnarray}\label{6.39}
&&E\bigl[L\bigl(\tilde{\theta}
_{(\cdot)}^{[2]}(\delta),\theta_{(\cdot)}\bigr)\bigr]
\nonumber
\\[-8pt]
\\[-8pt]
\nonumber
&&\quad \le E\bigl[L\bigl(\tilde{\theta} _{(\cdot)}^{[1]},\theta_{(\cdot)}\bigr)\bigr]\quad \mbox{for all }\gamma\le\delta\le 1.
\end{eqnarray}
Noting that $\mathop{\lim} _{m \to \infty} [(m - 1)^{2}/(m +
1)^{2}] = 1$, it follows that (\ref{6.39}) holds asymptotically for all
$\gamma$, and the inequality $\gamma\le\delta\le 1$ implies less
shrinkage of the direct estimators toward the mean. In particular, the
optimal choice of $\delta$ for $\tilde{\theta}_{(\cdot)} ^{[2]}(\delta)$ satisfies
$\lim_{m \to \infty} \delta^{\mathrm{opt}} = \gamma^{1/2}$.

The results above assume general distributions $H$ and $G$. When these\vadjust{\goodbreak}
distributions are normal, then for
$m = 2$, $E[L(\tilde{\theta} _{(\cdot)}^{[3]},\theta_{(\cdot)})] \le E[L(\tilde{\theta} _{(\cdot)}^{[2]}(\delta),\theta_{(\cdot)})]$ for all $\delta$. A conjecture
supported by simulations is that this relationship holds also for $m >
2$. However, the simulations suggest that for sufficiently large $m$
(e.g.,\vspace*{1pt} \mbox{$m \ge 25$}), $\tilde{\theta}_{(\cdot)} ^{[3]}$ is efficiently replaced by
$\tilde{\theta}_{(\cdot)} ^{[2]}(\gamma^{1/2})$. The last two conclusions are
shown empirically to hold also in the case where $\sigma_{u}^{2}$ is
unknown and replaced by the MOM variance estimator.

\begin{Remark}\label{rem5}
 The problem of predicting the ordered means is
different from ranking them, one of the famous triple-goal estimation
objectives in SAE. The triple-goal estimation consists of producing
``good'' area specific estimates, ``good'' estimates of the
histogram (distribution) and ``good'' estimates of the ranks. See
\citet{Ra03} for discussion. Judkins and Liu (\citeyear{JuLi00}) considered another
related problem of estimating the range of the area means. The authors
show theoretically and by simulations that the range of the direct
estimators overestimates the true range, whereas the range of the
empirical Bayes estimators underestimates the true range, in line with
the discussion at the beginning of this section. The bias is much
reduced by use of a constrained empirical Bayes estimator. For the model
considered by \citet{MaRi10}, the constrained estimator is
obtained by replacing the shrinkage coefficient $\gamma =
\sigma_{u}^{2}(\sigma_{u}^{2} + \sigma_{e}^{2})^{ - 1}$ in (\ref{5.2}) by
$\tilde{\gamma} \cong\gamma^{ - 1/2}$, which again shrinkages less the
direct estimator.
\end{Remark}

\subsection{New Developments for Specific Applications}\label{sec6.8}

In this section I review two relatively new applications of SAE;
assessment of literacy and poverty mapping. The latter application, in
particular, received considerable attention in recent years.

\subsubsection*{Assessment of literacy}

 The notable feature of
assessing literacy from a literacy test is that the possible outcome is
either zero, indicating illiteracy, or a positive continuous score
measuring the level of literacy. Another example of this kind of data is the consumption of
illicit drugs, where the consumption is either zero or a continuous
measure. In both examples the zero scores are ``structural'' (true)
zeroes. The common models used for SAE are not applicable for this kind
of responses if the proportion of zeroes is high. Pfeffermann, Terryn and Moura
(\citeyear{PfTeMo08}) consider the estimation of the average literacy score and the
proportion of people with positive scores in districts and villages in
Cambodia, a~study sponsored by the UNESCO Institute for Statistics (UIS).
Denote\vadjust{\goodbreak} by $y_{ijk}$ the test score of adult $k$ from village $j$ of
district $i$ and by $r_{ijk}$ a set of covariates and district and
village random effects. The following relationship holds:
\begin{eqnarray}\label{6.40}
E(y_{ijk}|r_{ijk}) &=& E(y_{ijk}|r_{ijk},y_{ijk} > 0)\nonumber\\[-8pt]\\[-8pt]
&&{}\cdot\Pr(y_{ijk} >
0|r_{ijk}).\nonumber
\end{eqnarray}
The two parts in the right-hand side of (\ref{6.40}) are modeled as
$E[y_{ijk}|r_{ijk},y_{ijk} > 0] = \mathrm{x}'_{ijk}\beta + u_{i} +
v_{ij}$, where $(u_{i},v_{ij})$ are district and nested village random
effects, $\Pr(y_{ijk} > 0|r_{ij}) =
p_{ijk}$;  $\operatorname{logit}(p_{ijk}) = \gamma '\mathrm{z}_{ijk} +
u_{i}^{*} + v_{ij}^{*}$, where $\mathrm{z}_{ijk}$ defines a set of
covariates which may differ from $\mathrm{x}_{ijk}$ and
$(u_{i}^{*},v_{ij}^{*})$ are district and nested village random effects,
which are correlated respectively with $(u_{i},v_{ij})$. The village and
district predictors of the average score and the proportion of positive
scores are obtained by application of the Bayesian approach with
noninformative priors, using MCMC simulations. The use of the Bayesian
approach enables one to account for the correlations between the
respective random effects in the two models, which is not feasible when
fitting the two models separately. The area predictors are obtained by
imputing the responses for nonsampled individuals by sampling from
their posterior distribution, and adding the imputed responses to the observed
responses (when observations exist).

\begin{Remark}\label{rem6}
 Mohadjer et al. (\citeyear{MoRaLiKrKe07}) estimate the proportions
$\theta_{ij}$ of adults in the lowest level of literacy in counties and
states of the USA, by modeling the direct estimates $\tilde{p}_{ij}$
in county $j$ of state $i$ as $\tilde{p}_{ij} = \theta_{ij} +
\varepsilon_{ij}$, and modeling $\operatorname{logit}(\theta_{ij}) =
x'_{ij}\beta + u_{i} + v_{ij}$ with $u_{i}$ and $v_{ij}$ defining state
and county random effects. The state and county estimates are likewise
obtained by MCMC simulations with noninformative priors. Note that this
is not a two-part model.
\end{Remark}

\subsubsection*{Poverty mapping}
The estimation of poverty indicators
in small regions is of major interest in many countries across the world,
initiated and sponsored in many cases by the United Nations and the
World Bank. In a celebrated article (awarded by the Canadian Statistical
Society as the best paper published in 2010 in \textit{The Candian
Journal of Statistics}), Molina and Rao focus on estimation of area
means of nonlinear poverty measures called FGT defined as
\begin{eqnarray}\label{6.41}
F_{\alpha i} &=& \frac{1}{N_{i}}\sum_{j = 1}^{N_{i}} F_{\alpha
ij};\nonumber\\[-8pt]\\[-8pt]
 F_{\alpha ij} &=& \biggl( \frac{z - E_{ij}}{z} \biggr)^{\alpha}
\times{I}(E_{ij} < z),\nonumber\\
\eqntext{\alpha = 0,1,2,}
\end{eqnarray}
where $E_{ij}$ is a measure of welfare for unit $j$ in area $i$ such as
income or expenditure, $z$ is a poverty threshold under which a person
is considered ``poor'' (e.g., 60\% of the nation median income)
and ${I}( \cdot)$ is the indicator function. For $\alpha = 0$,
$F_{\alpha i}$ is the proportion under poverty. For $\alpha = 1$,
$F_{\alpha i}$ measures the ``poverty gap,'' and for $\alpha = 2$,
$F_{\alpha i}$ measures ``poverty severity.''

For $\alpha = 1,2$ it is practically impossible to assign a distribution
for the measures $F_{\alpha ij}$, and in order to estimate the means
$F_{\alpha i}$ in sampled and nonsampled areas, Molina and Rao (\citeyear{MoRa10})
assume the existence of a one-to-one transformation $y_{ij} = T(E_{ij})$
such that the transformed outcomes $y_{ij}$ satisfy the unit level model
(\ref{5.3}) with normal distribution of the random effects and the residuals.
Notice that $F_{\alpha ij} = [1 - \frac{1}{z}T^{ - 1}(y_{ij})]^{\alpha}
\times{I}[T^{ - 1}(y_{ij}) < z] = : h_{\alpha} (y_{ij})$. For
sampled units $j \in s_{i}$ $F_{\alpha ij}$ is known, and for the
nonsampled units $k \in r_{i}$, the missing measures are imputed by
the EBP $F_{\alpha ik}^{\mathrm{EBP}} = \hat{E}[h_{\alpha} (y_{ik})|y_{s}] =
\sum_{l = 1}^{L} h_{\alpha} (y_{ik}^{(l)})/L$ with large $L$, where
$y_{s}$ defines all the observed outcomes. The predictions
$y_{ik}^{(l)}$ are obtained by Monte Carlo simulation from the
conditional normal distribution of the unobserved outcomes given the
observed outcomes under the model (\ref{5.3}), using estimated parameters
$\hat{\psi} = (\hat{\beta '},\hat{\sigma} _{u}^{2},\hat{\sigma}
_{\varepsilon} ^{2})'$. The PMSE of the EBP $\hat{F}_{\alpha i}^{\mathrm{EBP}} =
[\sum_{j \in s_{i}} F_{\alpha ij} + \break\sum_{k \in r_{i}} F_{\alpha
ik}^{\mathrm{EBP}} ]/N_{i}$ is estimated similarly to the first step of the
double-bootstrap procedure described in Section~\ref{sec6.1}. Model- and
design-based simulations and application to a real data set from Spain
using the transformation $y_{ij} = \log(E_{ij})$ demonstrate good
performance of the area predictors and the PMSE estimators.

\begin{Remark}\label{rem7}
The World Bank (WB) is currently using a different
method, under which all the population values $y_{ij}$ are simulated from model
(\ref{5.3}) with estimated parameters (including for sampled units), but with
random effects for design clusters, which may be different from the
small areas. As discussed and illustrated by \citet{MoRa10}, the use of this
procedure means that all the areas are practically considered as
nonsampled, and the resulting predictors of the means $F_{\alpha
i}$ in (\ref{6.41}) are in fact synthetic predictors since the random effects and the
area means of the residuals cancel out over the $L$ simulated
populations. Simulation results in Molina and Rao (\citeyear{MoRa10}) show that
the WB method produces predictors with much larger PMSE than the PMSE of
the EBP predictors proposed by them.
\end{Remark}

\section{SAE under Informative Sampling and Nonresponse}\label{sec7}

All the studies reviewed in this paper assume, at least implicitly, that
the selection of areas that are sampled and the sampling designs within
the selected areas are noninformative, implying that the model assumed
for the population values applies also to the sample data with no
sampling bias. This, however, may not be the case, and as illustrated
in the literature, ignoring the effects of informative sampling may bias
the inference quite severely. A~similar problem is not missing at random (NMAR)
nonresponse under which the response probabilities depend on the
missing data, which again can bias the predictions if not accounted
for properly. These problems received attention under both the
frequentist and the Bayesian approaches.

\citet{PfSv07} consider the problem of informative
sampling of areas and within the areas. The basic idea in this article
is to fit a sample model to the observed data and then exploit the
relationship between the sample model, the population model and the
sample-complement model (the model holding for nonsampled units) in
order to obtain unbiased predictors for the means in sampled and
nonsampled areas.

Consider a two-stage sampling design by which $m$ out of $M$ areas are
selected in the first stage with probabilities $\pi_{i} = \Pr(i \in s)$,
and $n_{i}$ out of $N_{i}$ units are sampled from the $i$th selected
area with probabilities $\pi_{j|i} = \Pr(j \in s_{i}|i \in s)$. Denote
by ${I}_{i}$ and ${I}_{ij}$ the sample indicator variables
for the two stages of sampling and by $w_{i} = 1/\pi_{i}$ and $w_{j|i} =
1/\pi_{j|i}$ the first and second stage sampling weights. Suppose that
the first level area random effects $\{ u_{1},\ldots,u_{M}\}$ are generated
independently from a distribution with p.d.f. $f_{p}(u_{i})$, and
that for given $u_{i}$ the second level values $\{
y_{i1},\ldots,y_{iN_{i}}\}$ are generated independently from a distribution
with p.d.f. $f_{p}(y_{ij}|x_{ij},u_{i})$.
The conditional first-level \textit{sample p.d.f.} of $u_{i}$, that is, the
p.d.f. of $u_{i}$ for area $i \in s$ is
\begin{eqnarray}\label{7.1}
f_{s}(u_{i})&\stackrel{\mathrm{def}}{ =}&  f(u_{i}|{I}_{i} =
1)\nonumber\\
&=& \operatorname{Pr}({I}_{i} = 1| u_{i})f_{p}(u_{i})/\Pr({I}_{i} = 1)\\
& =&
E_{s}(w_{i})f_{p}(u_{i})/E_{s}(w_{i}| u_{i}).\nonumber
\end{eqnarray}
The conditional first-level \textit{sample-complement} \textit{p.d.f.} of~$u_{i}$,
that is, the p.d.f. for area $i \notin s$ is
\begin{eqnarray}\label{7.2}
f_{c}(u_{i})&\stackrel{\mathrm{def}}{ =}&  f(u_{i}|{I}_{i} = 0)\nonumber\\[-8pt]\\[-8pt]
&=& \operatorname{Pr}({I}_{i} = 0| u_{i})f_{p}(u_{i})/\Pr({I}_{i} = 0).\nonumber
\end{eqnarray}
Note that the \textit{population}, \textit{sample} and
\textit{sample-comple\-ment} p.d.f.s are the same if
$\Pr({I}_{i} = 1|u_{i}) = \Pr({I}_{i} = 1)$, in which case
the area selection is \textit{noninformative}. Similar relationships
hold between the sample p.d.f., population p.d.f. and sample-complement p.d.f. of the outcomes $y_{ij}$ within the selected areas, for given
values of the random effects.

\citet{PfSv07} illustrate their approach by assuming
that the \textit{sample model} is the unit-level model (\ref{5.3}) with normal
random effects and residuals, and that the sampling weights within the
selected areas have sample model expectations,
\begin{eqnarray}\label{7.3}
&&E_{si}(w_{j|i}|\mathrm{x}_{ij},y_{ij},u_{i},{I}_{i} = 1)\nonumber\\
 &&\quad=
E_{si}(w_{j|i}|\mathrm{x}_{ij},y_{ij},{I}_{i} = 1)\\
 &&\quad=
k_{i}\exp(a'\mathrm{x}_{ij} + by_{ij}),\nonumber
\end{eqnarray}
where $k_{i} = N_{i}(n_{i})^{ - 1}\sum_{j = 1}^{N_{i}} \exp( - a'x_{ij} -
by_{ij}) /N_{i}$,\break and $a$ and $b$ are fixed constants. No model is
assumed for the relationship between the area selection probabilities
and the area means. The authors show that under this model and for given
parameters $\{ \beta',b,\sigma_{u}^{2},\sigma_{\varepsilon} ^{2}\}$, the
true mean $\bar{Y}_{i}$ in sampled area $i$ can be predicted as
\begin{eqnarray}\label{7.4}
\hspace*{25pt}\hat{\bar{Y}}_{i} &=& E_{p}(\bar{Y}_{i}|D_{s},{I}_{i} =
1)\hspace*{-25pt}\nonumber\\
&=&\frac{1}{N_{i}}\{ (N_{i} - n_{i})\hat{\theta} _{i} + n_{i}[\bar{y}_{i}
+ (\bar{X}_{i} - \mathrm{\bar{x}}_{i})'\beta ]\\
&&\hspace*{102pt}{}+ (N_{i} -
n_{i})b\sigma_{e}^{2}\},\nonumber
\end{eqnarray}
where $D_{s}$ represents all the known data and $\hat{\theta} _{i} =
\hat{u}_{i} + \bar{X}_{i}\beta$ is the optimal predictor of the sample
model mean $\theta_{i} = \bar{X}'_{i}\beta + u_{i}$. The last term in
(\ref{7.4}) corrects for the sample selection effect, that is, the difference
between the sample-complement expectation and the sample expectation in
sampled areas.

The mean $\bar{Y}_{k}$ of area $k$ not in the sample can be predicted as
\begin{eqnarray}\label{7.5}
&&\hat{E}_{p}(\bar{Y}_{k}|D_{s},{I}_{k} = 0)\nonumber\\
&&\quad=\bar{X}'_{k}\beta + b\sigma_{e}^{2}\\
&&\qquad{}+\biggl[\sum_{i \in s} (w_{i} - 1)
\hat{u}_{i}\Big/\sum_{i \in s} (w_{i} - 1)\biggr].\nonumber
\end{eqnarray}
The last term of (\ref{7.5}) corrects for the fact that the mean of the random
effects in areas outside the sample is different from zero under
informative selection of the areas. The authors develop test procedures
for testing the informativeness of the sample selection and a
bootstrap procedure for estimating the PMSE of the empirical predictors
obtained by substituting the unknown model parameters by sample
estimates. The method is applied for predicting the mean body mass index
(BMI) in counties of the USA using data from the third national health
and nutrition examination survey (NHANES
III).

Malec, Davis and Cao (\citeyear{MaDaCa99}, hereafter MDC) and Nandram and Choi (\citeyear{NaCh10}, hereafter
NC) likewise consider the estimation of county level BMI statistics from
NHANES III, with both articles accounting for within-area informative
sampling in a similar manner, and the latter article accounting, in
addition, for informative nonresponse. Another difference between the
two articles is that MDC consider binary population outcomes
(overweight/normal status), with logistic probabilities that contain
fixed and multivariate random area effects, whereas NC assume a
log-normal distribution for the continuous BMI measurement, with linear
spline regressions containing fixed and random area effects defining
the means. In order to account for sampling effects, both articles
assume that each sampled unit represents $K - 1$ other units (not sampled) within a
specified group (cluster) of units, with unit $j$ selected with
probability $\pi_{(j)}^{*}$ that can take one of the $G$ observed values\vspace*{1pt}
$\pi_{g}^{*}$, $g = 1,\ldots,G$ in that group.
The groups are defined by county and demographic characteristics.
Specifically, let $\delta_{j}
= 1\ (0)$ if unit $j$ is sampled (not sampled). The MDC model for a given group assumes
\begin{eqnarray}\label{7.6}
\hspace*{15pt}&&\delta_{j}|K,\pi_{(j)}^{*}\stackrel{\mathrm{ind}}{\sim}
\operatorname{Bernoulli}\bigl(\pi_{(j)}^{*}\bigr),\quad j =
1,\ldots,K;\hspace*{-15pt}\nonumber\\
&&\Pr\bigl(\pi_{(j)}^{*} = \pi_{g}^{*}|\theta_{gy},y_{j} = y\bigr) = \theta_{gy},\nonumber\\\\[-18pt]
&\eqntext{y= 0,1; g = 1,\ldots,G,} \\
&&\Pr(y_{j} = y|p) = p^{y}(1 - p)^{1 - y},\nonumber\\
&\eqntext{0 \le p
\le 1; p(K) = 1.}
\end{eqnarray}
It follows that
\begin{eqnarray}\label{7.7}
\hspace*{14pt}&&\mathrm{P} \bigl(\delta_{j} = 1,y_{j} =
y,\pi_{(j)}^{*} = \pi_{g}^{*},\{ \delta_{k} = 0\} _{k \ne j}|\theta,p\bigr)\hspace*{-14pt}\nonumber\\[-8pt]\\[-8pt]
&&\quad\propto\frac{p^{y}(1 - p)^{1 - y}}{\sum_{g = 1}^{G} \pi _{g}^{*}\sum_{y
= 0}^{1} \theta _{gy}p^{y}(1 - p)^{1 - y}}.\nonumber
\end{eqnarray}
MDC show that the MLE of $\theta_{gy}$ is $\hat{\theta} _{gy} =
(\tau_{gy}/\pi_{g}^{*})/\allowbreak\sum_{g* = 1}^{G} (\tau_{g*y}/\pi_{g*}^{*})$
where $\tau_{gy}$ is the sample frequency of $\pi_{g}^{*}$ in the group
for units with overweight status $y$. They plug the estimate into (\ref{7.7})
and then into the full likelihood that includes also the distribution of random effects contained in a logit model for
$p$.

NC generalize model (\ref{7.6}) by allowing the outcome to be continuous,
assuming\vspace*{1pt} $\Pr(\pi_{(j)}^{*} = \pi_{g}^{*}|\break\theta_{g}(y),y) =
\theta_{g}(y)$, $- \infty < y < \infty$ where $\theta_{g}(y) =
\theta_{gl}$ for $a_{l - 1} < y < a_{l}$, and replacing the Bernoulli
distribution for $y$ by a continuous p.d.f. To account for informative
nonresponse, the authors assume that the response probabilities
$p_{ij}^{r}$ are logistic with $\operatorname{logit}(p_{ij}^{r}) = v_{0i} +
v_{1i}y_{ij}$, where $\{ (v_{0i},v_{1i})\}$ is another set of random
effects having a bivariate normal distribution.

\begin{Remark}\label{rem8}
 As the notation suggests, both MDC and NC use the
full Bayesian approach with appropriate prior distributions to obtain
the small area predictors under the respective models. See the articles for
details. The authors do not consider informative sampling of the areas.
\end{Remark}

I conclude this section by describing an article by \citet{Zh09}, which
uses a very different model from the other models considered in the
present paper. The article considers the estimation of small area
compositions in the presence of NMAR nonresponse. Compositions
are the counts or proportions in categories of a categorical variable
such as types of households, and estimates of the compositions are
required for every area. Zhang deals with this problem by assuming that
the generalized SPREE model (GSPREE) developed in \citet{ZhCh04} holds for the complete data (with no missingness). In order to
account for the nonresponse, Zhang assumes that the probability to
respond is logistic, with a fixed composition effect $\xi_{c}$ and a
random area effect $b_{a}$ as the explanatory variables. (Same
probability for all the units in a given cell defined by
area $\times$ category.) The model depends therefore on two sets of random
effects, one set for the underlying complete data, with a vector of
correlated multivariate normal composition effects in each area defining
the GSPREE model, and the other set for the response probabilities. \citet{Zh09} estimates
the small area compositions under the extended\break GSPREE
using the EM algorithm, and estimates the PMSE
under the model, accounting for the fixed and random effects estimation.
The approach is applied to a real data set from Norway.

\section{Model Selection and Checking}\label{sec8}

Model selection and checking is one of the major problems in SAE because
the models usually contain unobservable random effects, with limited or
no information on their distribution. Notice that classical model
selection criteria such as the AIC do not apply straightforwardly to
mixed models because they use the likelihood, which requires
specification of the distribution of the random effects, and because of
difficulties in determining the effective number of parameters. In what follows I
review several recent studies devoted to model selection and validation
from both a frequentist and Bayesian perspective. These should be
considered as supplements to ``ordinary'' checking procedures
based on graphical displays, significance testing, sensitivity of the
computed predictors and their PMSEs to the choice of the likelihood and
the prior distributions, and comparison of the model-dependent
predictors with the corresponding model free direct estimators in
sampled areas. Such model evaluation procedures can be found in almost
every article on SAE; see, for example, Mohadjer et al. (\citeyear{MoRaLiKrKe07}) and
Nandram and Choi (\citeyear{NaCh10}) for recent diverse applications.

\citet{VaBl05} study the use of the AIC assuming model
(\ref{6.1}) with $\operatorname{Var}(u_{i}) = Q$,\break $\operatorname{Var}(e_{i}) = \sigma^{2}I_{n_{i}}$. The
authors distinguish between inference on the marginal model with focus
on the fixed effects, and inference on the model operating in the small
areas with the associated vector random effects $u_{i}$. For the first
case, the model can be written as a regression model with correlated
residuals: $y_{i} = X_{i}\beta + v_{i}$; $v_{i} = Z_{i}u_{i} + e_{i}\sim
N(0,Z_{i}QZ'_{i} + \sigma^{2}I_{n_{i}})$. For this case, the classical
(marginal) AIC, $\mathrm{mAIC} = \break- 2\log g(y|\hat{\psi} _{\mathrm{MLE}}) + 2P$ applies,
where $y$ is the vector of all the observations, $g(y|\hat{\psi}
_{\mathrm{MLE}})$ is the marginal likelihood evaluated at the MLE of $\psi$, the
vector containing $\beta$, $\sigma^{2}$ and the unknown elements of $Q$
and \mbox{$P = \dim(\psi)$}. \citet{Gu06} validates by simulations that one can
use also in this case the mAIC with $\hat{\psi} _{\mathrm{REML}}$, despite the
use of different fixed effects design matrices under different models.

For the case where the focus is the model operating at the small areas,
\citet{VaBl05} propose using a conditional AIC, which, for a
given likelihood $g(y|\psi,u)$, is defined as
\begin{eqnarray}\label{8.1}
\mathrm{cAIC} &=& - 2\log g(y|\hat{\psi} _{\mathrm{MLE}},\hat{u}) + 2P^*;\nonumber\\[-8pt]\\[-8pt]
 P^* &=&
\frac{n(n - k - 1)(\rho + 1) + n(k + 1)}{(n - k)(n - k - 2)},\nonumber
\end{eqnarray}
where $k$ is the number of covariates, $\hat{u} = E(u|\hat{\psi}
_{\mathrm{MLE}},y)$ is the EBP of $u$ and $\rho = \operatorname{tr}(H)$ with $H$ defining the
matrix mapping the observed vector $y$ into the fitted vector $\hat{y} =
X\hat{\beta} + Z\hat{u}$, such that $\hat{y} = Hy$. Notice that under
this definition of the cAIC, the $u_{i}$s are additional parameters. A conditional
AIC for the case where $\psi$ is estimated by REML is also developed.
The article contains theoretical results on properties of the cAIC and
empirical results illustrating its good performance. The use of (\ref{8.1}) is
not restricted to mixed linear models with normal distributions of the
error terms, and it can be used to select the design matrices $X_{i}$
and $Z_{i}$.

Pan and Lin (\citeyear{PaLi05}) propose alternative goodness-of-fit test statistics
for the GLMM, based on estimated cumulative sums of residuals. Utilizing
the notation for model (\ref{6.1}), the GLMM assumes the existence of a
one-to-one link function $g( \cdot)$, satisfying $g[E(y_{ij}|u_{i})] =
\mathrm{x}'_{ij}\beta + \mathrm{z}'_{ij}u_{i}$, where $\mathrm{x}_{ij}$
and $\mathrm{z}_{ij}$ are the rows of the matrices $X_{i}$ and $Z_{i}$
corresponding to unit $(i,j) \in s_{i}$. The unconditional predictor of
$y_{ij}$ is $m_{ij}(\psi) = E(y_{ij}) = E_{u_{i}}[g^{ -
1}(\mathrm{x}'_{ij}\beta + \mathrm{z}'_{ij}u_{i})]$, which is estimated by
$m_{ij}(\hat{\psi} )$. The estimated model residuals are therefore
$e_{ij} = y_{ij} - m_{ij}(\hat{\psi} )$, and they are computed by
numerical integration. The authors consider two statistics based on
the distributions of aggregates of the residuals,
\begin{eqnarray}\label{8.2}
W(\mathrm{x}) &=& n^{ - 1/2}\sum_{i = 1}^{m} \sum_{j = 1}^{n_{i}}
I(\mathrm{x}_{ij} \le\mathrm{x}) e_{ij},\nonumber\\[-8pt]\\[-8pt]
 W_{g}(r) &=& n^{ - 1/2}\sum_{i =
1}^{m} \sum_{j = 1}^{n_{i}} I(\hat{m}_{ij} \le r)
e_{ij},\nonumber
\end{eqnarray}
where $I(\mathrm{x}_{ij} \le\mathrm{x}) = \prod_{l = 1}^{k} I(x_{ijl}
\le x_{l})$. In particular, for~testing the functional form of the $l$th
covariate, one~may consider the process $W_{l}(x) = n^{ - 1/2}\cdot\allowbreak \sum_{i =
1}^{m} \sum_{j = 1}^{n_{i}} I(\mathrm{x}_{ijl} \le x) e_{ij}$, which is
a special case\break of~$W(\mathrm{x})$. The authors develop a simple
approximation for the null distribution of $W_{l}(x)$ and use it for
visual inspection by plotting the observed values against realizations
from the null distributions for different values of $x$, and for a
formal test defined by the supremum $S_{l} = \sup_{x}|W_{l}(x)|$. The
statistic $S_{l}$ is used for testing the functional form of the
deterministic part of the model. To test the appropriateness of the link
function, the authors follow similar steps, using the statistics
$W_{g}(r)$ for visual inspection and $S_{g} = \sup_{r}|W_{g}(r)|$ for
formal testing. As discussed in the article, although different tests
are proposed for different parts of the model, each test actually checks
the entire model, including the assumptions regarding the random
components.

The goodness-of-fit tests considered so far assume a given structure of
the random effects, but are random effects\vadjust{\goodbreak} actually needed in a SAE
model applied to a given data set? Datta, Hall and Mandal (\citeyear{DaHaMa11}) show that if in
fact the random effects are not needed and are removed from the model,
it improves the precision of point and interval estimators. The
authors assume the availability of $k$ covariates $\mathrm{x}_{i} =
(x_{1i},\ldots,x_{ki})$, $i = 1,\ldots,m$ (viewed random for the theoretical
developments) and weighted area-level means $\bar{y}_{i} = \sum_{j =
1}^{n_{i}} w_{ij}y_{ij}$; $\sum_{j = 1}^{n_{i}} w_{ij} = 1$ of the outcome
with known weights and known sums $W_{ir} = \sum_{j = 1}^{n_{i}}
w_{ij}^{r} $, $r = 2,\ldots,q$, $q \le k$. The weights $w_{ij}$ are used for
generating new area level means from bootstrap samples, and the sums
$W_{ir}$ are used for estimating model parameters by constructing
appropriate estimating equations.

In order to test for the presence of random effects, the authors propose
the test statistic
\begin{equation}\label{8.3}
\quad T = \sum_{i = 1}^{m} [W_{i2}\lambda_{2}(\mathrm{x}_{i},\hat{\psi}
)]^{ - 1}[\bar{y}_{i} - \lambda_{1}(\mathrm{x}_{i},\hat{\psi} )]^{2},
\end{equation}
where $\lambda_{l}(\mathrm{x}_{i},\hat{\psi} )$, $l = 1,2$ define the
conditional mean and residual variance of $y|\mathrm{x}$ under the
reduced model of no random effects, with estimated (remaining)
parameters $\hat{\psi}$. Critical values of the distribution of
$T$ under the null hypothesis of no random effects are obtained by
generating bootstrap samples with new outcomes from the conditional
distribution of $y|\mathrm{x};\hat{\psi}$ for given (original)
covariates and weights, and computing the test statistic for each
sample. Empirical results indicate good powers of the proposed procedure
and reduction in PMSE when the null hypothesis is not rejected. The
procedure is applicable to very general models.

Jiang et al. (\citeyear{JiRaGuNg08}) propose a class of strategies for mixed model
selection called \textit{fence methods}, which apply to LMM and GLMM.
The strategies involve a procedure to isolate a subgroup of correct
models, and then select the optimal model from this subgroup according
to some criterion. Let $Q_{M} = Q_{M}(y,\psi_{M})$ define a measure of
``lack of fit'' of a candidate model $M$ with parameters
$\psi_{M}$, such that $E(Q_{M})$ is minimized when $M$ is the true
model. Examples of $Q_{M}$ are minus the loglikelihood or the residual
sum of squares. Define $\hat{Q}_{M} = Q_{M}(y,\hat{\psi} _{M}) =
\inf_{\psi _{M} \in \Psi _{M}}Q_{M}(y,\psi_{M})$, and let $\tilde{M}
\in\mathrm{M}$ be such that $Q_{\tilde{M}} = \min_{M \in
\mathrm{M}}\hat{Q}_{M}$ where $\mathrm{M}$ represents the set of candidate models.
It is shown that under certain conditions, $\tilde{M}$ is a correct model
with probability tending to one. In practice, there can be more than one
correct model and a second step of the proposed procedure is to select
an optimal model among models that are within a fence around
$Q_{\tilde{M}}$. Examples of optimality criteria\vadjust{\goodbreak} are minimal dimension
or minimum PMSE. The fence is defined as $\hat{Q}_{M}
\le\hat{Q}_{\tilde{M}} + c_{n}\hat{\sigma} _{M,\tilde{M}}$, where
$\hat{\sigma} _{M,\tilde{M}}$ is an estimate of the standard deviation
of $\hat{Q}_{M} - \hat{Q}_{\tilde{M}}$,  and $c_{n}$ is a tuning
coefficient that increases with the total sample size. Jiang et al.
(\citeyear{JiRaGuNg08}) discuss alternative ways of computing $\hat{\sigma}
_{M,\tilde{M}}$ and propose an adaptive procedure for choosing the
tuning coefficient. The procedure consists of parametric bootstrapping
new samples from the ``full'' model, computing for every candidate
model $M \in\mathrm{M}$ the empirical proportion $p^{*}(M,c_{n})$ that
it is selected by the fence method with a given $c_{n}$, computing
$p^{*}(c_{n}) = \max_{M \in \mathrm{M}}p^{*}(M,c_{n})$ and choosing
$c_{n}$ that maximizes $p^{*}(c_{n})$.

Jiang et al. (\citeyear{JiRaGuNg08}) apply the method for selecting the covariates in the
area-level model (\ref{5.1}) and the unit level model (\ref{5.3}). Jiang, Nguyen and Rao
(\citeyear{JiNgRa10}) apply the method for selecting nonparametric P-spline models of
the form (\ref{6.31}). Selecting a model in this case requires selecting the
degree of the spline $p$, the number of knots $K$ and a smoothing
parameter $\lambda$ used for estimation of the model parameters.

So far I have considered model selection and diagnostic procedures under
the frequentist approach, but sound model checking is obviously required
also under the Bayesian approach. Although this article is concerned
with new developments, it is worth starting with a simulation procedure
proposed by Dey et al. (\citeyear{DeGeSwVl98}) since it highlights a possible advantage
of the Bayesian approach in model checking. Let $d$ define a discrepancy
measure between the assumed model and the data, such as minus
the first-stage likelihood of a hierarchical model. Denote by $y_{\mathrm{obs}}$
the observed data and assume an \textit{informative prior}. The
procedure consists of generating a large number $R$ of new data sets
$y_{\mathrm{obs}}^{(r)}$, $r = 1,\ldots,R$ under the presumed model via Monte Carlo
simulations and comparing the posterior distribution of $d|y_{\mathrm{obs}}$ with
the distributions of $d|y_{\mathrm{obs}}^{(r)}$. Specifically, for each posterior
distribution $f(d|y_{\mathrm{obs}}^{(r)})$ compute the vector of quantiles
$q^{(r)} = q_{\alpha _{1}}^{(r)},\ldots,q_{\alpha _{Q}}^{(r)}$ (say
$\alpha_{1} = 0.025$, $\ldots$\,, $\alpha_{Q} = 0.975$), compute $\bar{q} =
\sum_{r = 1}^{R} q^{(r)}/R$ and the Euclidean distances between
$q^{(r)}$ and $\bar{q}$, and check\break whether the distance of the quantiles
of the distribution of $d|y_{\mathrm{obs}}$ from $\bar{q}$ is smaller or larger
than, say, the 95th percentile of the $R$ distances.

\begin{Remark}\label{rem9}
 The procedure is computationally intensive, and it
requires informative priors to allow generating new data sets, but it is
very flexible in terms of the models tested and the discrepancy
measure(s) used. A frequentist analog via parametric bootstrap would
require that the distribution of $d$ does not depend on the model
parameters, or that the sample sizes are sufficiently large to permit
ignoring parameter estimation.
\end{Remark}

\citet{BaCa07} investigate Bayes\-ian methods for
\textit{objective} model checking, which requires
\textit{noninformative priors} for the parameters $\psi$. The authors
assume a given diagnostic statistic $T$ (not a function of $\psi$) and
consider two ``surprise measures'' of conflict between the observed data
and the presumed model; the $p$-value $\Pr^{h( \cdot )}[T(y) \ge\break
t(y_{\mathrm{obs}})]$, and the relative predictive surprise $\mathrm{RPS} =
h[t(y_{\mathrm{obs}})]/\sup_{t}[h(t)]$, where $h(t)$ is some specified
distribution. Denote by $\theta$ the small area parameters. Writing
$f(y) = \int f(y|\theta)g(\theta)\,d\theta$, it is clear that defining
$h(t)$ requires integrating $\theta$ out of $f(y|\theta)$ with respect
to some distribution for $\theta$. The prior $g(\theta)$ cannot be used
since it is also improper and the authors consider three alternative
solutions: 1. Set the model hyper-parameters $\psi$ at their estimated
value and integrate with respect to $g(\theta |\hat{\psi} )$. This is
basically an application of empirical Bayes and $h^{\mathrm{EB}}(t) = \int
f(t|\theta)g(\theta |\hat{\psi} )\,d\theta$. 2. Integrate $\theta$ out by
use of the posterior distribution $g(\theta |y_{\mathrm{obs}})$. 3.~Noticing that
under the above two solutions, the data are used both for obtaining a
proper distribution for $\theta$ and for computing the statistic
$t(y_{\mathrm{obs}})$, the third solution removes the information in $t(y_{\mathrm{obs}})$
from $y_{\mathrm{obs}}$ by using the conditional likelihood
$f(y_{\mathrm{obs}}|t_{\mathrm{obs}},\theta)$. The resulting posterior distribution for
$\theta$ is then used to obtain the distribution $h(t)$, similarly to
the previous cases. The specified distribution $h(t)$ under all three
cases may not have a closed form, in which case it is approximated by
MCMC simulations. See the article for details and for illustrations of
the approach showing, in general, the best performance under the third
solution.

\citet{YaSe07} consider a specific model inadequacy, namely,
fitting models that do not account for all the hierarchical structure
present, and, like the last article, restrict to noninformative priors.
The authors consider two testing procedures, both based on the
predictive posterior distribution $f(\tilde{y}|y_{\mathrm{obs}}) = \int
f(\tilde{y}|\psi)p(\psi |y_{\mathrm{obs}})\,d\psi$, where $\tilde{y}$ and $y_{\mathrm{obs}}$
are assumed to be independent given $\psi$. The first procedure uses the
posterior predictive $p$-values, $p_{ij} = \Pr(\tilde{y}_{ij} \le
y_{ij}|y_{\mathrm{obs}})$. The second procedure uses the $p$-values of a diagnostic
statistic $t( \cdot)$ or a discrepancy measure $d( \cdot)$ (see above),
for example, the $p$-values $\Pr [t(\tilde{y}) \ge t(y_{\mathrm{obs}})|y_{\mathrm{obs}}]$.
The authors analyze the simple case of a balanced sample where the
fitted model is $y_{ij}|\mu,\phi\stackrel{\mathrm{i.i.d.}}{\sim}  N(\mu,\phi)$,
$i = 1,\ldots,m$, $j = 1,\ldots, n_{0}$. It is shown that if the model is correct,
then as $N = n_{0}m \to\infty$ the distributions of $y_{\mathrm{obs}}$ and
$\tilde{y}|y_{\mathrm{obs}}$ are the same, and the $p$-values $p_{ij}$ are
distributed uniformly, as revealed in a Q--Q plot. On the other hand, if
the true model is the two-level model
$y_{ij}|\theta_{i},\phi_{0}\stackrel{\mathrm{i.i.d.}}{\sim}N(\theta_{i},\phi_{0})$,
$\theta_{i}|\mu_{0},A_{0}\stackrel{\mathrm{i.i.d.}}{\sim}N(\mu_{0},A_{0})$,
then as $N \to\infty$ the mean and variance of the two models still
agree, but not the covariances, so that it is the ensemble of the
$p_{ij}$s or their Q--Q plot against the uniform distribution, but not
individual $p$-values, that permits distinguishing the two models. This,
however, is only effective if the intra-cluster correlation is
sufficiently high, and the number of areas sufficiently small. Similar
conclusions hold when comparing a two-stage hierarchical model with a
three-stage model, and when applying the second testing procedure with
the classical ANOVA $F$ test statistic as the diagnostic statistic, that
is, when computing $\Pr [F(\tilde{y}) \ge F(y_{\mathrm{obs}})|y_{\mathrm{obs}}]$.

\citet{YaSe10} consider a third procedure for detecting a
missing hierarchical structure, which uses Q--Q plots of the predictive
standardized residuals $r_{ij} = \frac{y_{ij} -
E(\tilde{y}_{ij}|y_{\mathrm{obs}})}{[\operatorname{Var}(\tilde{y}_{ij}|y_{\mathrm{obs}})]^{1/2}}$ against
the standard normal distribution. The conditions under which the
procedure performs well in detecting a misspecified hierarchy are the
same as above.

Finally, I like to mention two articles that in a certain way bridge
between the frequentist and Bayesian approaches for model selection. The
idea here is to set up a noninformative prior under the Bayesian
approach so that the resulting posterior small area predictors have
acceptable properties under the frequentist approach. This provides
frequentist validation to the Bayesian methodology, and the analyst may
then take advantage of the flexibility of Bayesian inference by drawing
observations from the posterior distribution of the area parameters. Both
articles consider the area-level model (\ref{5.1}), but the idea applies to
other models.

Datta, Rao and Smith (\citeyear{DaRaSm05}) assume a flat prior for $\beta$ and seek a prior
$p(\sigma_{u}^{2})$ satisfying $E(V_{i\mathrm{HB}}) = \operatorname{PMSE}[\hat{\theta}
_{i}(\hat{\sigma} _{u,\mathrm{RE}}^{2})] + o(m^{ - 1})$, where $V_{i\mathrm{HB}}
=\break
\operatorname{Var}(\theta_{i}|y_{\mathrm{obs}})$ is the posterior variance of $\theta_{i}$,
and\break
$\operatorname{PMSE}[\hat{\theta} _{i}(\hat{\sigma} _{u,\mathrm{RE}}^{2})]$ is the frequentist
PMSE of the\linebreak[4] EBLUP (or EB) when estimating $\sigma_{u}^{2}$ by REML. The\vadjust{\goodbreak}
expectation and PMSE are computed under the joint distribution of
$\theta$ and $y$ under the model. The unique prior satisfying this
requirement is shown to be
\begin{equation}\label{8.4}
\qquad p_{i}(\sigma_{u}^{2}) \propto(\sigma_{Di}^{2} +
\sigma_{u}^{2})^{2}\sum_{j = 1}^{m} [1/(\sigma_{Dj}^{2} + \sigma_{u}^{2}
)^{2}].
\end{equation}
The prior is area specific in the sense that different priors are
required for different areas.

\citet{GaLa08} extend the condition of Datta, Rao and Smith (\citeyear{DaRaSm05}) to a
weighted combination of the posterior expectations and the PMSEs, thus obtaining a single prior for all
the areas. The authors seek a prior which for a given set of weights $\{
\omega_{i}\}$ satisfies
\begin{eqnarray}\label{8.5}
&&\sum_{i = 1}^{m} \omega_{i} \{ E(V_{i\mathrm{HB}}) - \operatorname{PMSE}[\hat{\theta}
_{i}(\hat{\sigma} _{u,\mathrm{RE}}^{2})]\}\nonumber\\[-8pt]\\[-8pt]
&&\quad = o(1/m).\nonumber
\end{eqnarray}
The prior $p(\sigma_{u}^{2})$ satisfying (\ref{8.5}) is shown to be
\begin{eqnarray}\label{8.6}
p(\sigma_{u}^{2}) &\propto&\sum_{i = 1}^{m} [1/(\sigma_{Di}^{2} +
\sigma_{u}^{2} )^{2}]\nonumber\\[-8pt]\\[-8pt]
&&{}\bigg/\sum_{i = 1}^{m}
\omega_{i}[\sigma_{Di}^{2}/(\sigma_{Di}^{2} + \sigma_{u}^{2} )]^{2}.\nonumber
\end{eqnarray}
By appropriate choice of the weights $\{ \omega_{i}\}$, prior (\ref{8.6})
contains as special cases the flat prior $p(\sigma_{u}^{2}) =
U(0,\infty)$, the prior developed by Datta, Rao and Smith (\citeyear{DaRaSm05}) for a given
area and the average moment matching prior (obtained by setting
$\omega_{i} \equiv 1)$.

\section{Concluding Remarks}\label{sec9}

In this article I reviewed many new important developments in design-
and model-based SAE. These developments give analysts much richer and
more versatile tools for their applications. Which approach should one
follow in practice? Model-based predictors are generally more accurate
and, as discussed in Section~\ref{sec4.3}, the models permit predictions for
nonsampled areas for which no design-based theory exists. With
everything else that can be done under a model, much of which reviewed
in Sections~\ref{sec6}--\ref{sec8}, it seems to me that the choice between the two
approaches is clear-cut, unless the sample sizes in all the areas are
sufficiently large, although even in this case models have much more to
offer like, for example, in the case of measurement errors or
NMAR nonresponse. This is not to say that design-based\vadjust{\goodbreak}
estimators have no role in model-based prediction. To begin with, the
design-based estimators are often the input data for the model, as under
the area-level model. Design-based estimators can be used for assessing
the model-based predictors or for calibrating them via benchmarking, and
the sampling weights play an important role when accounting for
informative sampling.

Next is the question of whether to follow the Bayes\-ian approach (BA) or
the frequentist approach (FA). I~have to admit that before starting this
extensive review I was very much in favor of FA, but the BA has some
clear advantages. This is because one can generate as many observations
as desired from the posterior distributions of the area parameters, and
hence it is much more flexible in the kind of models and inference
possibilities that it can handle, for example, when the linking model
does not match the conditional sampling model (Remark~\ref{rem2}). Note also that
the computation of PMSE (Bayes risk) or credibility intervals under BA
does not rely on asymptotic properties. A common criticism of BA is that it
requires specification of prior distributions but as emphasized in
Section~\ref{sec8}, Bayesian models with proper, or improper priors can be tested
in a variety of ways. Another criticism is that the application of BA is
often very computation intensive and requires expert knowledge and
computing skills even with modern available software. While this
criticism may be correct (notably in my experience), the use of FA
methods when fitting the GLMM is also very computation intensive and
requires similar skills. Saying all this, it is quite obvious to me that
the use of FA will continue to be dominant for many years to come
because, except for few exceptions, official statistical bureaus are
very reluctant to use Bayesian methods.

Where do we go from here? Research on SAE continues all over the world,
both in terms of new theories and in applications to new intriguing
problems, and I~hope that this review will contribute to this research.
The new developments that I have reviewed are generally either under BA
or FA, and one possible direction that I~hope to see is to incorporate
the new developments under one approach into the other. For example, use
the EL approach under FA, use spline regressions under BA, account for
NMAR nonresponse in FA or produce poverty mapping with BA. Some
of these extensions will be simple; other may require more extensive
research, and some may not be feasible, but this will make it easier for
analysts to choose between the two approaches.\vadjust{\goodbreak}

\section*{Acknowledgment}

I am very grateful to three reviewers for very constructive comments
which enhanced the discussion and coverage of this review very
significantly.


\end{document}